\documentclass[modern,trackchanges]{aastex631}

\usepackage{amssymb, amsmath}
\usepackage{color}
\usepackage{comment}
\usepackage{morefloats}
\usepackage{hyperref}

\begin{document}

\title{A Survey for C\,{\sc ii} Emission-Line Stars in the Large Magellanic Cloud.\\
II. Final Results and the Origin of C\,{\sc ii} Emission in [WC] Spectra}
 
\author{Bruce Margon}
\affiliation{Department of Astronomy \& Astrophysics, University of California, Santa Cruz, 1156 High St., Santa Cruz, CA 95064, USA} \email{margon@ucsc.edu}

\author{Nidia Morrell}
\affiliation{Las Campanas Observatory, Carnegie Observatories, Casilla 601, La Serena, Chile}
\email{nmorrell@carnegiescience.edu}

\author{Philip Massey}
\affiliation{Lowell Observatory, 1400 W Mars Hill Road, Flagstaff, AZ, 86001, USA}
\affiliation{Department of Astronomy and Planetary Science, Northern Arizona University, Flagstaff, AZ, 86011-6010, USA}
\email{phil.massey@lowell.edu}

\author{Kathryn F. Neugent}
\altaffiliation{NASA Hubble Fellow}
\affiliation{Center for Astrophysics, Harvard \& Smithsonian, 60 Garden St., Cambridge, MA 02138, USA}
\affiliation{Lowell Observatory, 1400 W Mars Hill Road, Flagstaff, AZ, 86001, USA}
\email{Kathryn.Neugent@cfa.harvard.edu}

\author{Robert Williams}
\affiliation{Space Telescope Science Institute, 3700 San Martin Drive, Baltimore, MD 21218, USA}
\affiliation{Department of Astronomy \& Astrophysics, University of California, Santa Cruz, 1156 High St., Santa Cruz, CA 95064, USA} \email{wms@stsci.edu}

\begin{abstract}

We present the final results of an imaging and spectroscopic search for stars in the Large Magellanic Cloud (LMC) with C\,{\sc ii} $\lambda\lambda$7231, 7236 emission lines. The goal is to identify and study [WC11] stars, the coolest of the low-mass Wolf-Rayet sequence, and a subset of central stars of planetary nebulae where the C\,{\sc ii} lines are known to be especially prominent. A recent serendipitous discovery of an LMC [WC11] raised the possibility that these objects, although difficult to identify, might in fact be more common than previously believed. Several new members of this rare class have been found in this survey. It now seems clear, however, that a significant number of these stars are not hiding amongst the general [WC] population. We point out that the C\,{\sc ii} doublet intensity ratio observed in our spectra proves to neatly divide the objects into two distinct groups, with the C\,{\sc ii} emission likely originating from either the stellar wind or a surrounding nebula. The physics of the C\,{\sc ii} emission mechanism correctly explains this bifurcation. Spectral subtypes are suggested for most of the objects. The numerous spectroscopic clues now available for these objects should facilitate future detailed modeling.

\end{abstract}




\section{Introduction}

The transition between the asymptotic giant branch (AGB) and the white
dwarf stages is one of the least understood phases of stellar evolution
despite recent progress in modeling  (e.g.,
\citealt{2016A&A...588A..25M}). Some unknown fraction of these
transition stars are believed to undergo a late thermal pulse, losing
their hydrogen-rich envelopes, mixing evolved products from the core,
and developing optically thick stellar winds. This phase is very short,
possibly only few hundred years, based on the rarity of these objects
(e.g., 
\citealt{1999A&A...349L...5H,herwig,2010aste.book.....A,2016A&A...588A..25M})
and results in a spectrum that superficially resembles a Wolf-Rayet (WR)
star within a planetary nebula (PN).

The modifier “superficial” is appropriate.  These objects resemble the
WR stars via their blue color and prominent broad C and He
emission lines; in both cases these lines are formed in an optically thick,
outflowing stellar wind.  However, WR stars are massive stars that have been stripped of their hydrogen-rich outer layers by stellar winds and/or binary interactions, while these low-mass analogs are the remnants of much lower mass stars that have undergone late thermal pulses.  Also unlike the WR stars, there is a carbon-rich, but no nitrogen-rich, sequence for these low-mass counterparts. Following a suggestion by \citet{vanderHucht81}, square brackets are normally used to denote this spectral sequence as [WC], and thus distinguish them from the classic, Population I high-mass WC stars. Assignment of the spectral subclass is somewhat author-dependent, but the coolest end, where the C\,{\sc ii} emission is the most prominent, is often denoted as [WC11] \citep{Crowther, Acker}.

\citet{Margon20b} (hereafter Paper~I) describe a number of issues regarding the coolest [WC] stars that are difficult to resolve given the current, quite small sample. The Large Magellanic Cloud (LMC) is well-suited for a search to enlarge this sample: with Galactic examples typically at $M_V\sim -3$, these stars will have $V\lesssim16$ in the LMC, and the modest LMC sky area is amenable to a comprehensive imaging survey. Paper~I describes such a survey, based on narrowband imaging to identify C\,{\sc ii} $\lambda\lambda$7231, 7236 emitters, and includes a small spectroscopic reconnaissance of the candidates, identifying several new, interesting examples. Here we report spectroscopy of the entire group of high priority candidates, which has identified multiple interesting stars.

\section{Spectroscopy of C II  Emission Candidates}
\subsection{Observations}
Paper~I describes in detail how candidates were selected, so that material will not be repeated here. Figure~2 of that paper displays the spatial distribution of the candidates within the LMC; they are distributed relatively at random throughout the bar and the outskirts of the galaxy. That publication also reports a preliminary spectroscopic reconnaissance of a small number of the candidates, using the MagE Spectrograph on the 6.5m Baade Magellan telescope of the Las Campanas Observatory. The observations described here used this same configuration, yielding complete wavelength coverage from 3150\,\AA, near the atmospheric cutoff, up to $\sim$10,000 \AA\, at intermediate resolution ($\lambda$/$\Delta\lambda\sim$ 4100). 

In addition to the observations described in Paper~I, during 2020 November and December we obtained further spectra of the most promising  C\,{\sc ii} emission candidates, each selected in the imaging survey described in Paper~I. There were a total of 28 stars observed, with typical exposure times of 600 -- 1200~s. In cases where particularly strong emission lines overwhelmed the continuum, multiple exposures of different integration times were used to obtain good signal-to-noise throughout the spectrum. A few objects were observed on multiple nights to search for evidence of radial velocity or spectral variability. 

\subsection{Spectral Classification}

All of these stars appear in one or more previous astrometric and/or photometric surveys, so there is no evidence that any of the objects are transient, at least not on a timescale of several decades. Many of the sample, however, have no further information in the literature and thus have been effectively anonymous. All of the objects observed prove likely LMC members, displaying multiple emission and absorption line redshifts of $\sim$\,6\,\AA, consistent with the LMC systemic radial velocity of +280~km~s$^{-1}$ \citep{Richter}.  Similarly, their Gaia DR3 \citep{gaia1,gaia3} proper motions are small (a few mas~yr$^{-1}$ or less), and their parallaxes too small to be meaningful.

Nine of the objects in our sample display spectra with prominent C\,{\sc ii} $\lambda\lambda$7231, 7236 emission. After the LMC [WC11] star J06081992-7157373 was discovered serendipitously by \citet{Margon20a}, \citet{Williams} presented line identifications for a high signal to noise spectrum of that star. The most prominent features those authors identify are more than 150 C\,{\sc ii} emission lines. Numerous Balmer and He\,{\sc i} emission features, many with P~Cyg profiles, are also present, as well as many narrow metallic absorption lines.  Thus, that paper is an excellent source of line identifications for similar C\,{\sc ii} emitters.

The basic properties of the subset of spectra displaying C\,{\sc ii} emission are given in Table~1; information on the remaining spectra appears in the Appendix.  Four of these C\,{\sc ii} emission stars are discussed in Paper~I, but are also included here, to provide a complete, homogeneous sample. All the objects discussed here are in the 2MASS catalog \citep{2MASS}, so we adopt that nomenclature. The coordinates inferred from the 2MASS object name may differ very slightly from Gaia DR3 \citep{gaia3}, typically by no more than a few tenths of an arcsecond (due both to the different precision of the two surveys, as well as the IAU nomenclature guidelines which truncate rather than round off the least significant position digits). However, in all cases the positions derived from the table entries are quite sufficient to unambiguously identify the stars for further observations, or to cross correlate with other catalogs.

We suggest spectral subclasses for each star, relying upon the criteria proposed by \citet{Crowther}. They argue for uniform and quantitative criteria for both [WC] and WC stars, and their primary subtype classifier is based upon the ratio of the equivalent widths of  C\,{\sc iv} $\lambda\lambda$5801, 5812 to that of C\,{\sc iii} $\lambda5696$. Secondary criteria include the full-width-at-half-maximum of the C\,{\sc iii} line, and, for the earliest types, the strength of O\,{\sc v} $\lambda5590$.

Our classification scheme diverges from the \citet{Crowther} criteria, however, at the very coolest end, as exemplified in the spectrum of the bright Galactic prototype cool [WC] star,  CPD\,$-56^\circ$\,8032 (= PN G332.9--09.9, = V837~Ara). We have a very high quality spectrum of this star obtained in 2018 February, also with MagE, and displayed in \citet{Margon20a}. Although \citet{Crowther} consider this star to be a [WC10], we detect no C\,{\sc iv} $\lambda\lambda$5801, 5812 emission, and thus classify this object as [WC11].  The structure in the nearby continuum is complex, but it appears that this doublet is weakly present, but in {\it absorption}. We see a nearly identical structure in our spectrum of J06081992-7157373, and indeed \citet{Williams} list that doublet as detected in absorption. We also have a high-resolution spectrum \hbox{($R\sim30,000$)} available of the latter star, taken with MIKE, the Magellan Inamori Kyocera Echelle Spectrograph, on the 6.5-m Clay telescope, which clearly confirms this. Thus, for the [WC11] class we suggest the criteria that C\,{\sc ii} is the dominant C ion, and that C\,{\sc iv} $\lambda\lambda$5801, 5812 is either missing or in absorption. Although \citet{Acker} also suggest the [WC11] classification when the C\,{\sc iv} doublet is in absorption, they also classified their spectrum of CPD\,$-56^\circ$\,8032 as [WC10], despite the apparent lack of C\,{\sc iv} emission in their displayed data. As with many issues in spectral classification, however, a divergence of $\pm1$ in spectral subclass is not uncommon, and likely should not be a matter of concern.

\subsection{Comments on Individual Objects}

\begin{figure}
\epsscale{1.2}
\plotone{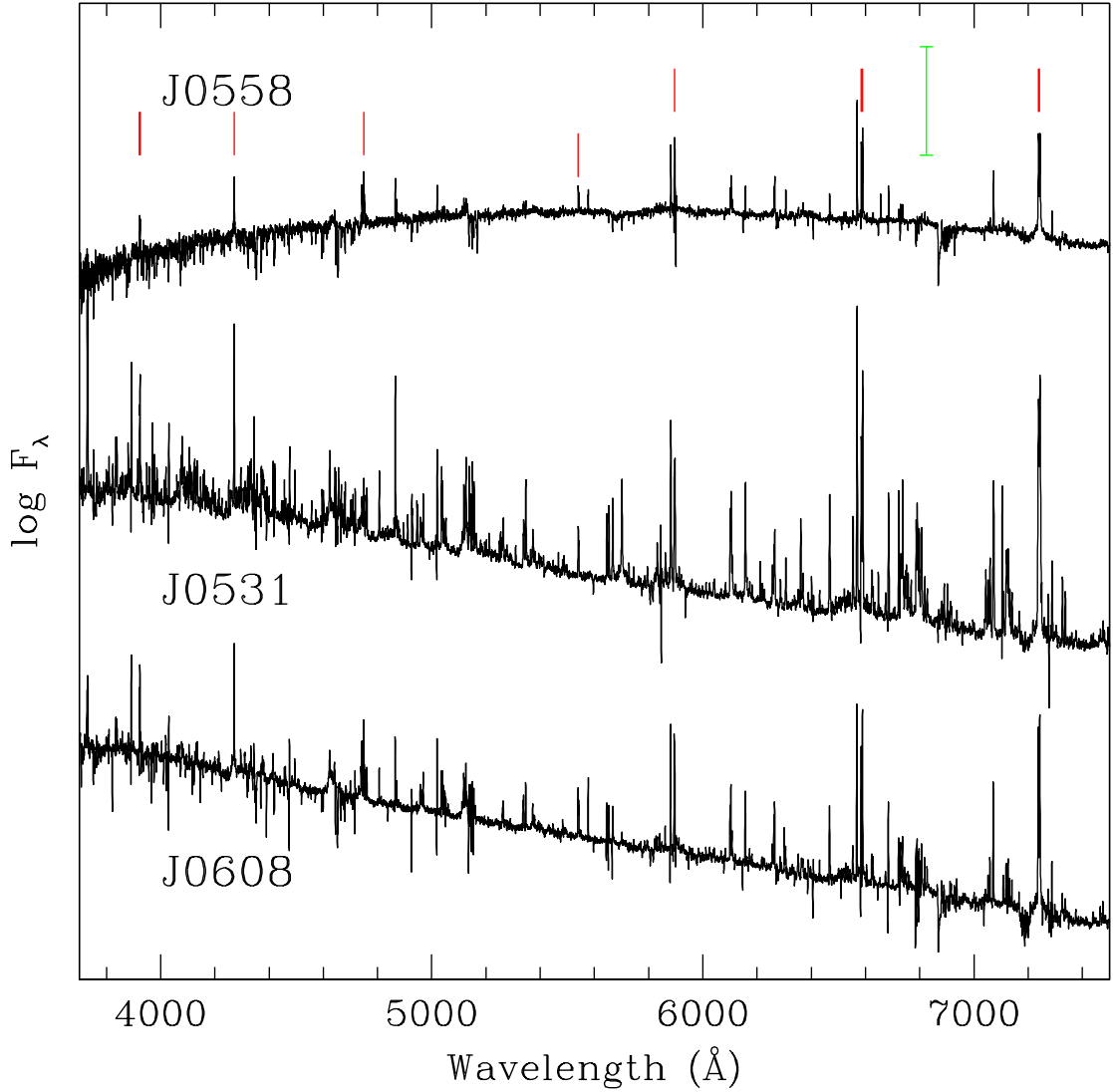}
\caption{\label{fig:late} The spectra of the two LMC [WC11] stars found as part of
our survey are compared to that of J0608. A comparison of the spectrum of J0608 and that of the archetype [WC11] star CPD\,$-56^\circ$\,8032 can be found in \citet{Margon20a}. The multiple, intense C\,{\sc ii} emission lines are apparent in all three objects, with prominent examples marked with vertical red bars. An exhaustive list of line identifications is given by \citet{Williams}. The spectra are plotted as log flux, in order to facilitate comparison of emission-line strengths with the continuum; the vertical green bar indicates a change of 0.5 dex. Although J0531 and J0608 have nearly identical SEDs, the spectrum of J0558 is considerably depressed in the blue and UV. These data, as well as those in subsequent figures, are not corrected for telluric absorption. For clarity, star names are abbreviated from Table~1. The reduced FITS-format spectra are provided here as Data Behind Figure.
}
\end{figure}

\begin{figure}
\plotone{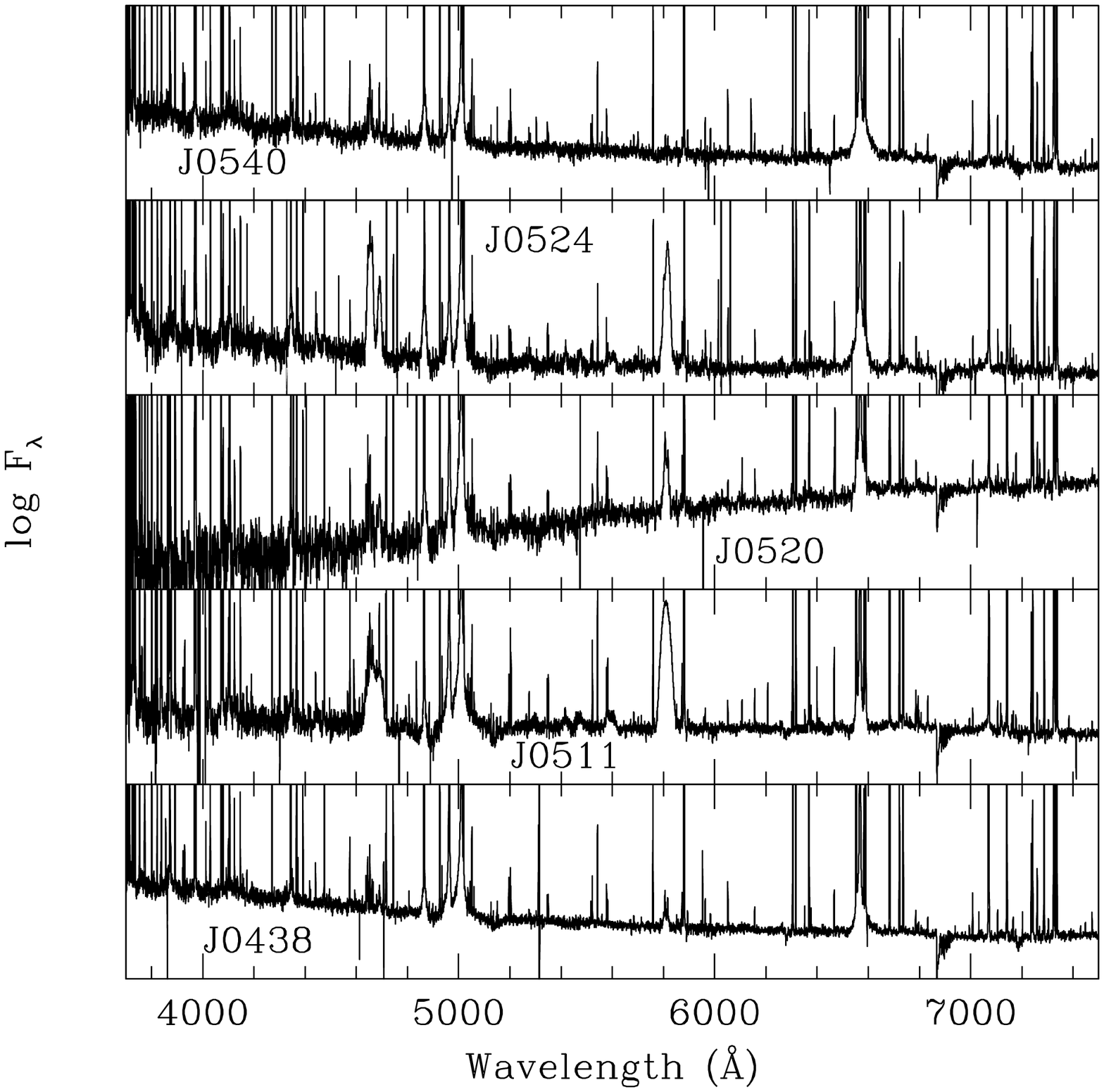}
\plottwo{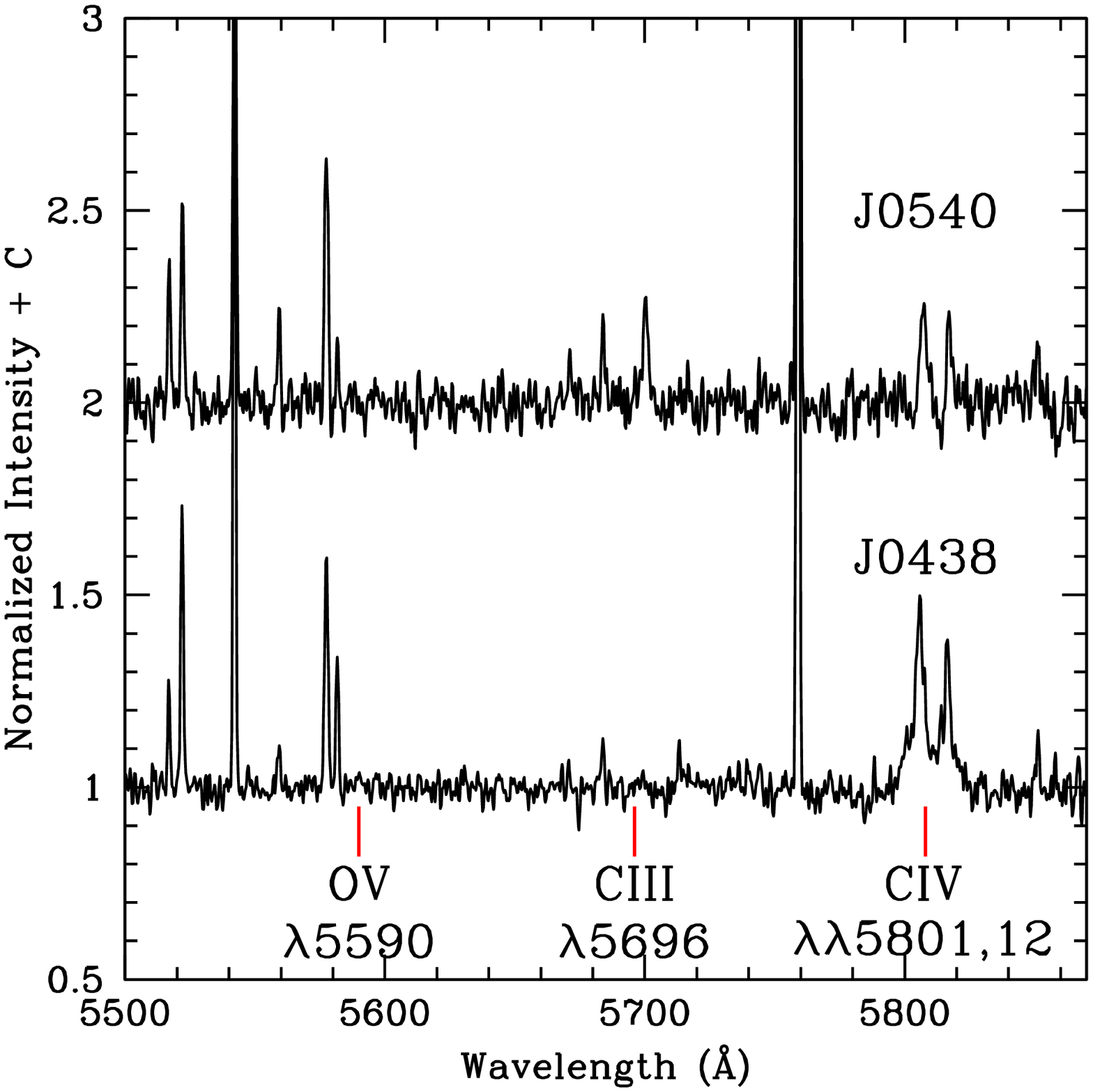}{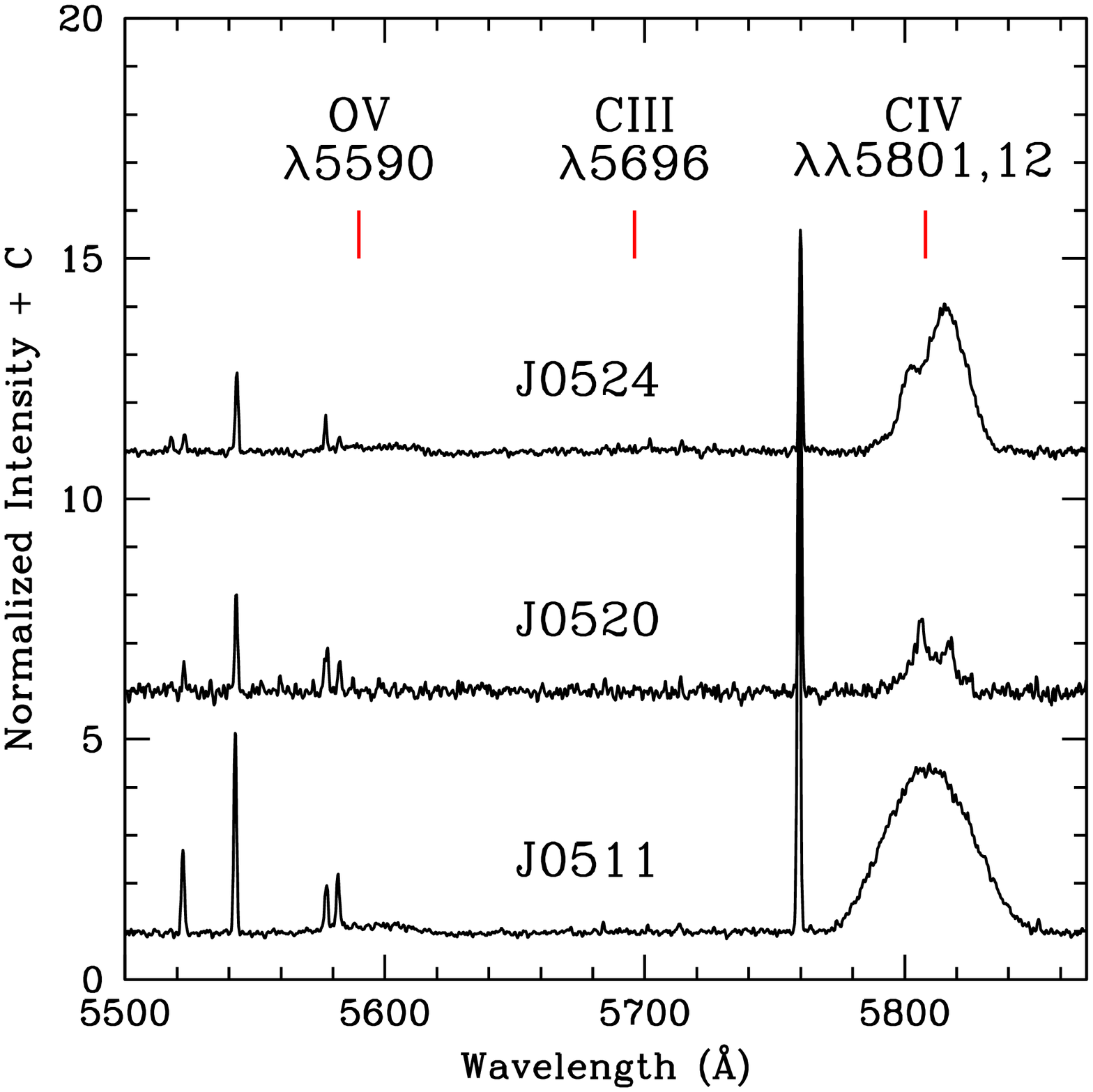}
\caption{\label{fig:early} Early-type LMC [WC] stars in our spectroscopic sample.
 {\it Upper:} Each panel is 1~dex high in $\Delta \log F_\lambda$ to facilitate comparison.  The spectra are strongly dominated by nebular emission in all cases.  Note, however, the presence of broad C\,{\sc iv} $\lambda \lambda$5801, 5812 emission in all but J0540. The reduced FITS-format spectra are provided here as Data Behind Figure. {\it Lower:} normalized intensities of the spectral region containing the O\,{\sc v}, C\,{\sc iii}, and C\,{\sc iv} classification features.
 }
\end{figure}

We present below suggested spectral subclasses as well as comments on each object in Table~1. In Figures~\ref{fig:late} and \ref{fig:early} we present the spectra of the late ([WC11]) and ``early" type [WC4-8] stars, respectively, with classifications made as described below. The individual spectra are also available in FITS format as ``data behind figures" in the online Journal.
 
\bigskip

{\it J04383478-7036434}: This is the central star of the well known PN WS1 \citep{Westerlund64} = LMC SMP1 \citep{SMP}. Despite a half-century of study, there is little in the literature on the stellar spectrum, which is overwhelmed by the nebular emission. \citet{Meat91} do note the presence of a weak  C\,{\sc ii} $\lambda 4267$ emission line. 

In our spectrum it is clear that numerous, prominent and narrow C\,{\sc ii} emission lines are present.  The data show no C\,{\sc iii} $\lambda$5696, and we set a maximum emission equivalent width (EW) value of -0.1\,\AA.  The components of the C\,{\sc iv} $\lambda\lambda5801, 5812$ doublet are well separated, but situated on a broad, underlying emission.  The EW of the entire feature is -4.4\,\AA. A broad feature (thus definitely not nebular) is visible at He\,{\sc ii} $\lambda4686$, with EW -1.5\,\AA\ and full-width-at-half-maximum (FWHM)  5.5\,\AA\ 
(350~km~s$^{-1}$). The C\,{\sc iv} to C\,{\sc iii} ratio is $>$40, which would classify this star as [WC4], but we note that the structure of C\,{\sc iv} is unlike other [WC4] stars. We would expect the C\,{\sc iv} $\lambda4650$ feature to be broad and strong; instead, the region is complex with multiple non-stellar emission components.

 \bigskip
 
 {\it J05073893-6826061}: Reported in Paper~I, where it is termed 233-1 (using our original survey nomenclature), this object has a particularly complex spectrum,  displayed in detail as Figure~7 of Paper~I, consisting of very strong C\,{\sc ii} $\lambda\lambda$7231, 7236 emission, but no other prominent emission, plus hundreds of narrow absorption lines of He, Si, and O. In addition to the discussions of the discovery spectrum in Paper~I and \citet{Margon21}, we now have four spectra spaced by weeks and months in the year 2020. Some of these have revealed weak emission of other C\,{\sc ii} transitions, particularly $\lambda4267$, and also possible H$\alpha$ absorption. While we suspect these detections represent subtle but genuine spectral variability, this interpretation must remain preliminary pending further confirmation. The collection of spectra show no radial velocity variability to a limit of $<$10~km~s$^{-1}$, making a binary explanation unlikely. There is still little clarity on the nature of this interesting star.
 
 \bigskip

{\it J05112369-7001573}: The central star of this PN (WS15, LMC SMP38) is classified as [WC4] by \citet{Pena}. Those authors tabulate weak C\,{\sc ii} $\lambda 4267$ emission. Our well-exposed data show multiple, strong C\,{\sc ii} lines throughout the spectrum, as well as strong, broad C\,{\sc iv} $\lambda\lambda$5801, 5812 emission
(EW -135\,\AA, FWHM 38\,\AA). There is no detectable C\,{\sc iii} $\lambda5696$.  Broad C\,{\sc iii}, C\,{\sc iv} $\lambda4650$ is also present, as is O\,{\sc v} $\lambda\lambda5572-98$. The C\,{\sc iii}, C\,{\sc iv} $\lambda$4650 and He\,{\sc ii} $\lambda$4686 lines are blended together, as is typical in the earliest [WC] stars, although there are numerous additional narrow emission lines superposed on this feature, likely nebular in origin. 

Given the strength of C\,{\sc iv} $\lambda5801, 5812$ and lack of C\,{\sc iii} $\lambda5696$ we would also term this star  [WC4]. However, the C\,{\sc iv} line width is more in keeping with [WC5], and O\,{\sc v} $\lambda 5590$ is weak rather than ``moderate” in strength; the latter is a secondary criterion for classification of WC4 stars \citep{Crowther, vanderHucht2001}.  In contrast to the broad spectrum, C\,{\sc ii} $\lambda 4267$ line has EW -20\,\AA\ and is unresolved at our resolution.  We classify this star [WC5], and note that the broad-line component is much more typical of a [WC] than several of the other stars discussed here (e.g., J04383478-7036434 and J05205243-7009354).

\bigskip

{\it J05205243-7009354}: This is the central star of the known PN LMC SMP51, with spectrophotometry of the nebula presented by \citet{Monk}. Our spectrum of the central star shows weak C\,{\sc iv} $\lambda \lambda$5801, 5812 emission which is double peaked, with each component having a FWHM of 7-9~\AA\ ($\sim400$~km~s$^{-1}$). The total EW of the feature is -20~\AA.  Once again, there is no C\,{\sc iii} present, again with upper limit -0.1~\AA.  We term this star [WC4--5], but acknowledge that the C\,{\sc iv} profile is highly unusual. Broad He\,{\sc ii} $\lambda$4686 emission is present, with EW -8~\AA\, and FWHM 830~km~s$^{-1}$.  Only nebular lines are present where we would expect to find the C\,{\sc iii}/C\,{\sc iv} $\lambda$4650 feature of an early [WC] star. The C\,{\sc ii} $\lambda 4267$ line and the components of C\,{\sc ii} $\lambda \lambda$7231, 7236 doublet are unresolved at our spectral resolution.  The EWs would also suggest [WC4] classification, but the lack of the 4650~\AA\ feature is indeed unusual.
\bigskip

{\it J05242076-7005015}: This star is reported in Paper~I, where it is termed 152-1, and more details and references may be found there. It hosts a known PN, LMC SMP58. C\,{\sc iv} $\lambda\lambda$5801, 5812 is present as two broad, blended components, with EW -77~\AA\ and FWHM 1300~km~s$^{-1}$.  The O\,{\sc v} $\lambda5590$ feature is also present, with EW -8~\AA.  Again, C\,{\sc iii} $\lambda$5696 is absent.  Both the C\,{\sc iii}/C\,{\sc iv} $\lambda$4650 and He\,{\sc ii} $\lambda 4686$ features are present, with EWs -67~\AA\ and -20~\AA, respectively. The latter has FWHM 1100~km~s$^{-1}$. The C\,{\sc ii} line at $\lambda 4267$ and the $\lambda \lambda$7231, 7236 doublet have EWs -13~\AA\ and -23~\AA, respectively. We thus classify this star [WC5], slightly cooler than the [WC4] suggested by \citet{Monk}.

In addition to the spectrum discussed in Paper~I, we now have multiple spectra spaced throughout the year 2020 and can limit any radial velocity variations to \hbox{$<$10~km~s$^{-1}$}.  As discussed further below, the lack of C\,{\sc iii} despite the presence of C\,{\sc iv} and C\,{\sc ii} drove us to suggest that this was a [WC4]+[WC11] binary.  The discovery of four other similar ``hybrid" stars in this current study, combined with the lack of radial velocity variations, is incompatible with this interpretation.  We discuss alternative explanations in \S3 below.

\bigskip

{\it J05312172-7017394}: This is a [WC11] unreported prior to our survey, and described in detail in Paper~I, where it is denoted as 153-1. The spectrum is similar to the LMC [WC11] prototype, J06081992-7157373 \citep{Margon20a, Williams}. There is no PN reported in the literature, but multiple forbidden lines in our spectrum indicate that a low-density region is certainly present.
\bigskip
 
{\it J05403079-6617374}: This is the central star of a known PN, LMC SMP85. The nebula is well-studied spectroscopically, and the central star successfully modeled as luminous, low-mass, and carbon-rich \citep{Dopita, Mashburn}. The UV spectrum is known to display C\,{\sc iii},  C\,{\sc iv}, and numerous other high excitation emission lines \citep{Herald}.

This spectrum is unique in our sample.  The C\,{\sc iv} $\lambda\lambda$5801, 5812 is double-peaked with FWHMs of only 120--180~km~s$^{-1}$ each, and a combined EW of only -1.5~\AA.  C\,{\sc iii} $\lambda5696$ is present in emission, the only star in our sample to show it, with EW -0.7~\AA\ and FWHM 130~km~s$^{-1}$.  C\,{\sc iii}/C\,{\sc iv} $\lambda$4650 and He\,{\sc ii} $\lambda4686$ are present, with EW -5~\AA\ and -3~\AA, respectively.  C\,{\sc ii} $\lambda$4267 and the $\lambda\lambda$7231, 7236 doublet have EW -6~\AA\ and -17~\AA, respectively, and their components are spectrally unresolved.  By the EWs of C\,{\sc iii} and C\,{\sc iv}, we would term this star [WC7], while the C\,{\sc iv} to C\,{\sc ii} ratio would suggest [WC9]. We tentatively term this a [WC8] central star. 

\bigskip 
 
{\it J05582596-6944257}: This star was advanced as a possible [WC12] by \citet{vanaarle}, and our spectrum confirms their classification as a late-type [WC]. However, this object deserves further study. Although the C\,{\sc ii} emission lines are numerous, they are weaker compared to the continuum than that of the other two late-type [WC] stars we show in Figure~\ref{fig:late}. \citet{hrivnak} comment that the object is ``likely [a] hot PPN/young PN."  \citet{Mats} present an infrared spectrum which displays polycyclic aromatic hydrocarbons (PAHs), as well as additional complex and unidentified features; they classify the object as a ``carbon rich post-AGB star." 

No nebula has been reported, and, unlike the other late [WC] stars discussed here, no [O\,{\sc ii}], [O\,{\sc iii}], or [S\,{\sc ii}] emission appears in our spectrum, which is well exposed. H$\alpha$ and H$\beta$ are in emission, although not obviously present in the spectrum presented by \citet{vanaarle}; the higher order Balmer lines are absent. There is substantially less UV-excess than in most of the LMC late [WC] stars; the $(B-V)$ value is also considerably redder than that of the other two [WC11] stars, and it is clear from Figure~\ref{fig:late} that the overall spectral energy distribution (SED) fails to match that of the other two. Based on the criteria of \S2.2, we classify this object as [WC11], slightly earlier than \citet{vanaarle}.

\bigskip

{\it J06081992-7157373}:   This is the initial, serendipitously identified LMC [WC11] \citep{Margon20a}, which motivated the current survey. A comprehensive discussion of the spectrum is given in that reference and in \citet{Williams}, and as noted above, we classify this stars as [WC11].

\bigskip
\section{Discussion}
 
\subsection{Space Density and Luminosity of LMC Late [WC] Stars}
 
\citet{Margon20a} pointed out that their serendipitous discovery during an entirely unrelated program of a relatively rare object such as a [WC11] star was either extremely lucky, or might instead indicate that the number of these stars may have been significantly underestimated in the past.  Although we cannot claim that our current LMC survey is complete, the spectroscopy of the candidates reported here, revealing only a modest number of new late-type [WC] stars in the $\sim50$~deg$^2$ survey area, does not support the speculation that the LMC (and by analogy the Milky Way) contains very large numbers of undiscovered cool [WC] stars. 

 It is of interest to compare
the luminosity distribution of the coolest LMC [WC] stars with those in the Milky Way.  There will likely be a large bolometric correction for these stars, but without detailed modeling, we cannot estimate its size.  The broad-band colors of two of the [WC11] stars are quite similar, but the third (J05582596-6944257) is much redder.  An inspection of Figure~\ref{fig:late} shows that this is not due to emission contamination but rather the continuum shape itself. Our attempt to deredden the spectrum with the standard assumptions of $R_V=3.1$ and a CCM law \citep{CCM} fails to match the spectral energy distribution of the other two, suggesting that the reddening is circumstellar, with a non-standard reddening law.  For now, we compare the absolute visual magnitudes, adopting a constant extinction correction of $A_V=0.4$ (\citealt{LGGSII} and references therein) and assuming an LMC true distance modulus of $(m-M)=18.47$~mag (49.4~kpc) \citep{Piet}.   We find that the $V$-band luminosity of the LMC late [WC] stars in our sample lies in the perhaps surprisingly narrow range of $-3.5\lesssim M_V\lesssim -2.5$. Applying an additional reddening correction to J05582596-6944257 to match the $B-V$ values of the other two [WC11] stars would extend this range, not tighten it.  We note that this absolute magnitude range is actually not very different from that of classical Population I WC stars, which are typically $M_V\sim-4.5$ \citep{2020MNRAS.493.1512R}.

Of course, as the stars in the sample are selected to have similar spectral features, namely the presence of C\,{\sc ii} in emission, similar luminosity might be expected. The lack of less luminous LMC examples is likely not entirely due to the sensitivity cutoff of our survey. J05312172-7017394, for example, is one of the two apparently faintest objects in Table~1, yet appears in our C\,{\sc ii} $\lambda\lambda$7231, 7236 narrow-band filter photometry with a $62\sigma$ excess over background \citep{Margon20a}.

These luminosity values can be compared with the prototypical Galactic late [WC] stars, CPD\,$-56^\circ$\,8032, M4-18, and He2-113, first proposed as a related group by \citet{Webster}. They have been studied for decades, and now all of them also have accurate {\it Gaia} parallaxes. We again assume the bolometric correction is similar for all of the objects, and correct for interstellar extinction, significant for these Milky Way objects, using the data of \citet{Schlegel}.  Given the various approximations, the resulting $V$-band luminosities are in reasonable agreement with those of our LMC sample; there is no evidence for any systematic offset of luminosities between  [WC] stars in the two galaxies.

\subsection{Origin of the C II emission}

Our spectrum of J05242076-7005015 reported in Paper~I presented us with a true quandary: the star had strong C\,{\sc ii} $\lambda \lambda$7231, 7236 emission but also strong C\,{\sc iv} $\lambda \lambda$5801, 5812 emission.  C\,{\sc iii} $\lambda 5696$, however, was completely absent.  How could a star have both C\,{\sc iv} {\it and} C\,{\sc ii} emission but no C\,{\sc iii}?  We were driven to the unlikely explanation that the star was a binary, consisting of a [WC11] and [WC4] pair. Given the evolutionary challenges involved in producing such a system (e.g., \citealt{herwig, demarco02}), this explanation seemed problematic. Our scenario became increasingly implausible when we observed four additional such ``hybrid" objects.  As discussed earlier, repeated measurements of J05242076-7005015 show no sign of radial velocity variations, providing further evidence that our initial explanation was improbable. 

There are multiple possibilities for both the location and the physical mechanism generating the C\,{\sc ii} emission in these stars, and we briefly explore them here. As all of the Galactic late [WC] stars are also the central stars of PNe, and most of the stars in Table~1 not already known to host PNe do display forbidden emission lines, it is natural to consider a nebula as a source of the emission. At the LMC distance, however, typical PNe are not easily resolved in ground-based observations. HST imaging shows that, with a handful of exceptions, the LMC PNe are found to be $\lesssim$ 0\farcs5 in extent \citep{Stan, Shaw}. Those authors present images of four of the five previously known PNe in Table~1, and they are resolved at about that angular size. Therefore direct observational attribution of the C\,{\sc ii} to a nebula rather than the central star via spatially resolved spectroscopy is not easily feasible from the ground.

Another plausible point of origin is the stellar wind, known to be present in the Galactic examples of cool [WC] stars. In these objects, the nebula and central star are well separated, and extremely strong C\,{\sc ii} emission is observed and known to be stellar. Indeed, intense C\,{\sc ii} emission has been recognized for decades as the primary classification attribute of the late [WC] stars, with the stellar wind designated as the probable source \citep{DeMarcoI, DeMarcoII, DeMarcoIII}. There are detailed models for this scenario \citep{Leu1, Leu2, Leu3}.

To add to the complexity of the possibilities, \citet{Hajduk} discuss a Galactic case where they identify the $\lambda\lambda$7231, 7236 emission as both stellar and also nebular.

\subsection{Emission Mechanisms}

\begin{figure}
\plottwo{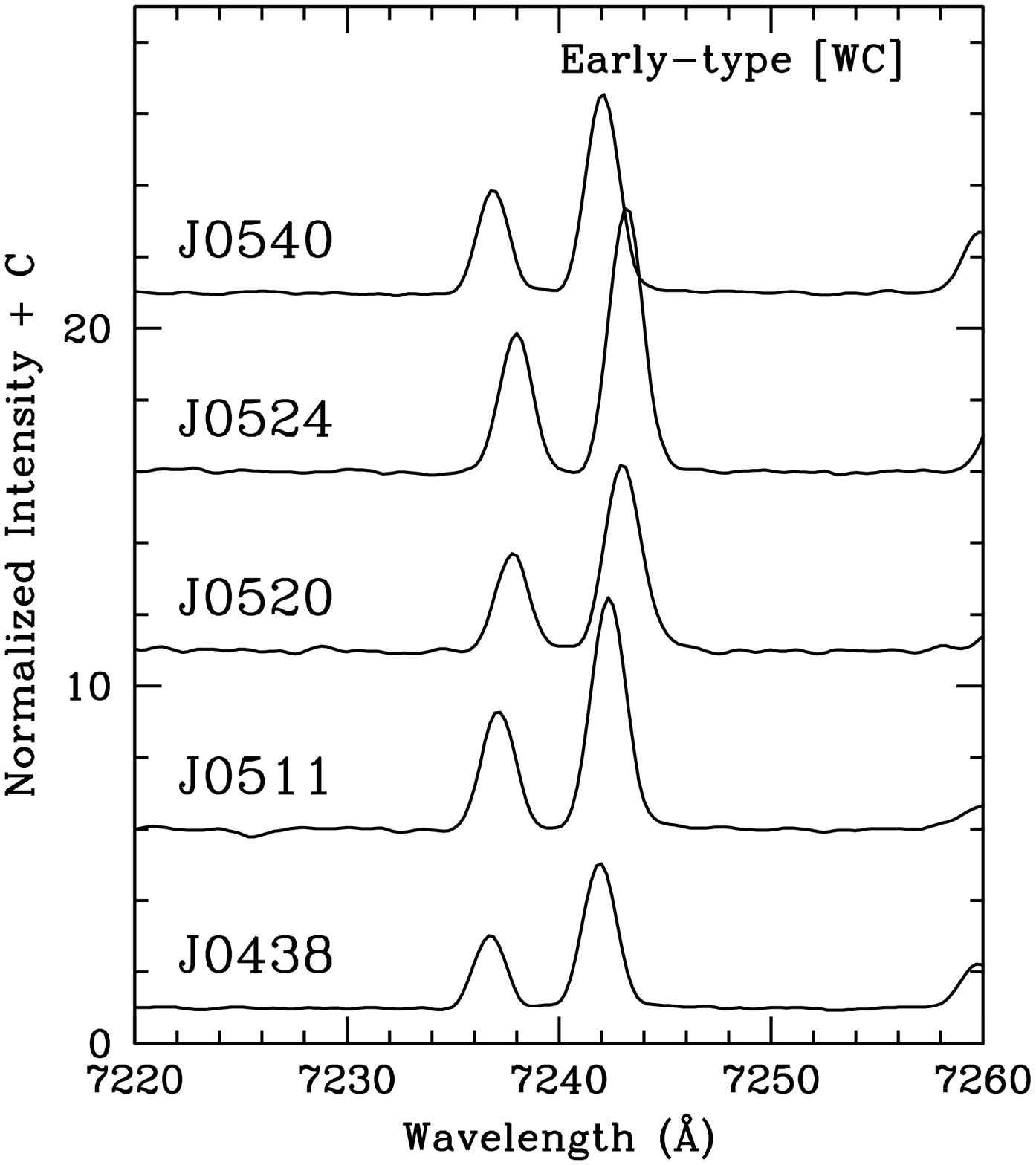}{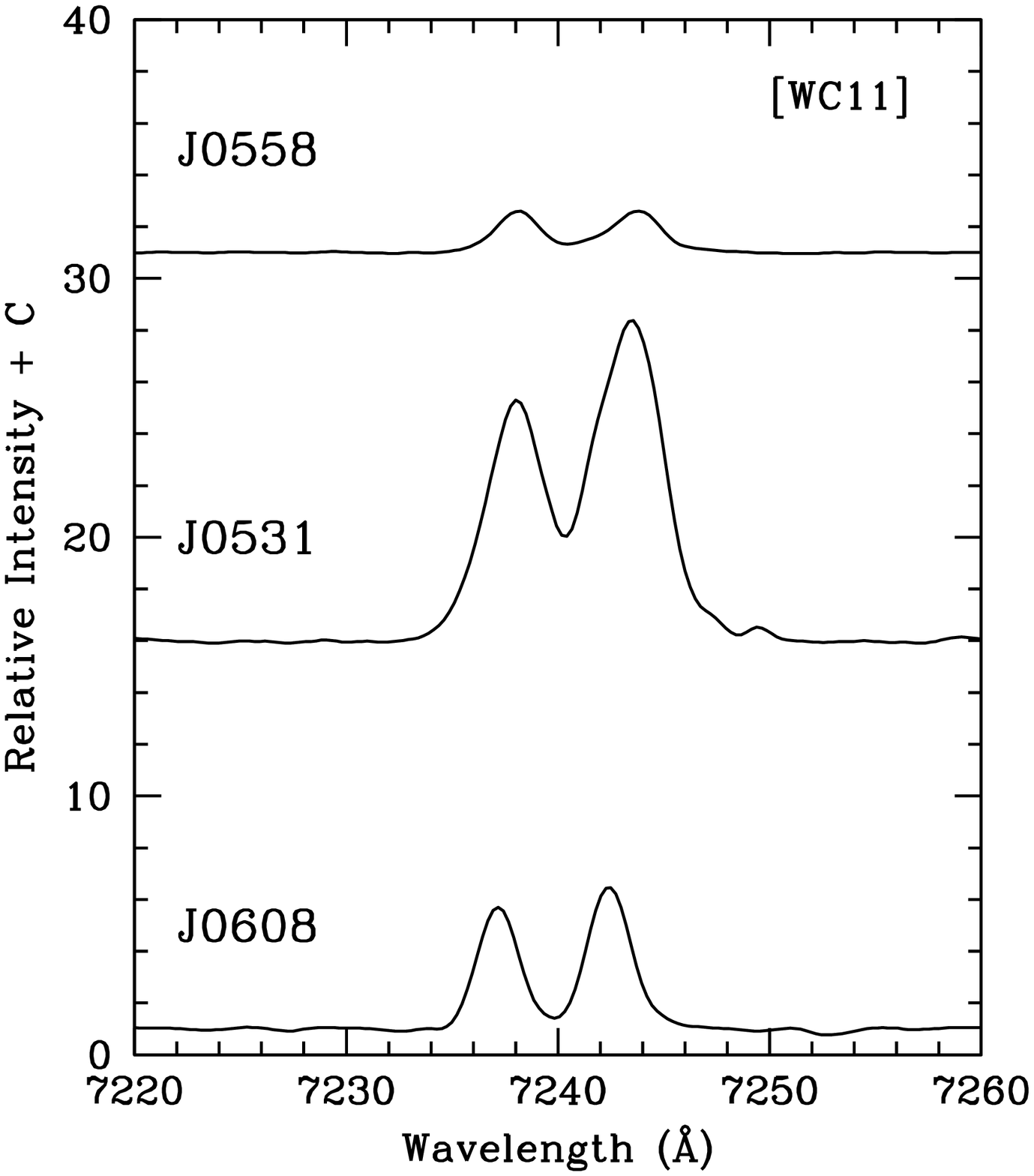}
\caption{\label{fig:CIILine} The relative intensities of the C\,{\sc ii} $\lambda \lambda$7231, 7236 doublet.  {\it Left:} line profiles of the C\,{\sc ii} for the early-type [WC] stars.  The two components are uneven, with the 7231\,\AA\ about half the strength of the 7236\,\AA\ component.  This is expected if the emission region is optically thin, as is the case if it is nebular in origin. {\it Right:} line profiles of the doublet for the [WC11] stars.  The two components are of nearly equal strength, as would be expected in the optically thick case, which would hold if the origin is from a stellar wind.
}
\end{figure}

We suggest that one can resolve the question of the origin of the C\,{\sc ii} emission by comparing the ratio of the intensities of individual lines in the C\,{\sc ii} multiplet to that expected from atomic physics. We include in Table 1 two key intensity ratios involving the C\,{\sc ii} doublet measured from our spectra, namely $\lambda7231/H\beta$, useful as an indicator of the overall strength of the C\,{\sc ii} emission, and also $\lambda$7231/$\lambda$7236, the ratio of the two strongest lines of the C\,{\sc ii} multiplet.  One notable result evident in these observed ratios is the bifurcation of the C\,{\sc ii} emitters into two quite distinct classes. Those objects with quite strong C\,{\sc ii} emission lines, $\lambda$7231/H$\beta \geq 1$, also display $\lambda$7231/$\lambda$7236 $\sim$1, while those with still prominent but weaker C\,{\sc ii} emission have the ratio of the C\,{\sc ii} doublet lines clustered very tightly near 0.5. This is readily apparent in Figure~\ref{fig:CIILine}.  We argue that this doublet ratio provides a reliable empirical separator of stellar versus nebular emission.

In the optically thin case (as when the origin is nebular), the intensity ratios of lines within a multiplet should be proportional to the ratios of their transition probabilities times the statistical weights of their upper levels, $g_J$ = 2J+1. The 7236 \AA\ component actually consists of two closely spaced lines, 7236.4~\AA\ and 7237.2~\AA, that are not resolved whenever the spectral resolving power is less than 6000 (i.e., 50 km s$^{-1}$),  and whose intensity is dominated by the 7236.4~\AA\ line.  For optically thin recombination lines, the expected line ratio of the 7231~\AA\ component to the unresolved 7236~\AA\ component is 0.51. This ratio is in fact exactly what we observe in the cases where there is strong nebulosity and the presence of C\,{\sc iv} but no C\,{\sc iii}. Thus in these cases, the C\,{\sc ii} emission is nebular in origin.

However, the expected ratio should be close to unity in the case of C\,{\sc ii} formed in a stellar wind, because the upper level of the $\lambda$7236 multiplet (of the three member lines) is the same upper level (3d $^2$D) as the strong C\,{\sc ii} resonance line $\lambda \lambda$687.1, 687.4, whose scattering of stellar continuum radiation populates its upper levels. That can cause the population of that $^2$D level to deviate from its otherwise 2J+1 statistical weight distribution among its two fine-structure states.  In short, in a nebula the resonance line does not play a strong role in competing with C$^{+2}$  electron recombination in populating the upper levels of the $\lambda$7236 multiplet because of the large dilution factor for the stellar continuum.  A wind, with a higher density and a stronger radiation field (where resonance scattering causes deviation of the population of the two $^2$D fine-structure states from their 2J+1 detailed balancing ratios)  yields a line ratio $\lambda$7236/$\lambda$7231 closer to unity. Therefore a line ratio near unity may be taken as evidence of a wind, with a strong continuum intensity near 687~\AA\ permeating the emitting gas. The latter is characteristic of the [WC11] class, whereas the origins of the C\,{\sc ii} emission in the [WC4–8] objects is nebular.

The fact that a substantial fraction of the stars in Table~1 exhibit nebular C\,{\sc ii} emission seems at first glance somewhat odd. In numerous well-studied Galactic PNe, where possible confusion of the spectrum of the central star and the surrounding nebulously is not an issue, strong nebular C\,{\sc ii} emission is not commonly reported. For example, although C\,{\sc ii} $\lambda 4267$ emission is often used in PNe abundance analyses \citep{Kwitter}, it is typically observed at intensity $\sim$$10^{-2}$ H$\beta$ \citep{Otsuka}, in approximate agreement with theoretical models, and far weaker than in about half of the spectra discussed here. The $\lambda7231$ line is typically weaker still, at $\sim$$10^{-3}$ H$\beta$ \citep{Zhang}. However, there are occasional examples of very metal-rich PNe where $\lambda7231$ is reported as strong as \hbox{$\sim$$10^{-2}$ H$\beta$}, comparable to values seen in Table~1 \citep{Liu}. This may provide an explanation for why, in the cases where the C\,{\sc ii} emission is nebular, our imaging survey isolates only a handful of PNe, despite the presence of $\sim$$10^3$ such objects in the LMC \citep{jacoby}: our survey may be selecting only the most metal-rich peak of the distribution.

\section{Summary}

Although the coolest [WC] stars are elusive to discover, the modest number of objects found in our survey implies that they are not a substantial fraction of the overall [WC] population. There is no evidence for any systematic differences between the LMC and the Galactic population of these objects. The prominent C\,{\sc ii} $\lambda\lambda$7231, 7236 emission can originate either in a nebula or a stellar wind. The doublet intensity ratio proves to be a convenient observational separator of the two cases, and thus may be useful in observations of galaxies substantially more distant than the LMC.

More definitive inferences on the nature of these objects will have to await analysis of each of the individual spectra with a detailed radiative transfer model such as \mbox{CMFGEN} \citep{Hillier98, Hillier01}, as pointed out by \citet{Williams}.

We thank the Mt.\ Cuba Astronomical Foundation for their generous support, which enabled purchase of the customized interference filters used for this survey. Useful correspondence with Anna O’Grady, John Hillier, and Micha\l~K. Szyma\'{n}ski is acknowledged. Comments by the anonymous referee helped to improve the manuscript. The observations reported here were possible thanks to the excellent assistance from the LCO technical and support staff.  This work was supported by the National Science Foundation (NSF) under AST-1612874 and through NASA Hubble Fellowship grant HST-HF2-51516. The Space Telescope Science Institute (STScI) is operated by AURA, Inc., under NASA contract NAS5-26555.

This work has made use of data from the European Space Agency (ESA) mission
{\it Gaia} (\url{https://www.cosmos.esa.int/gaia}), processed by the {\it Gaia}
Data Processing and Analysis Consortium (DPAC,
\url{https://www.cosmos.esa.int/web/gaia/dpac/consortium}). Funding for the DPAC
has been provided by national institutions, in particular the institutions participating in the {\it Gaia} Multilateral Agreement.  This research has made use of the VizieR catalogue access tool, CDS, Strasbourg, France (doi: 10.26093/cds/vizier).

\facilities{Magellan:Baade, Swope}

\eject

\begin{figure}
\epsscale{1.3}
\plotone{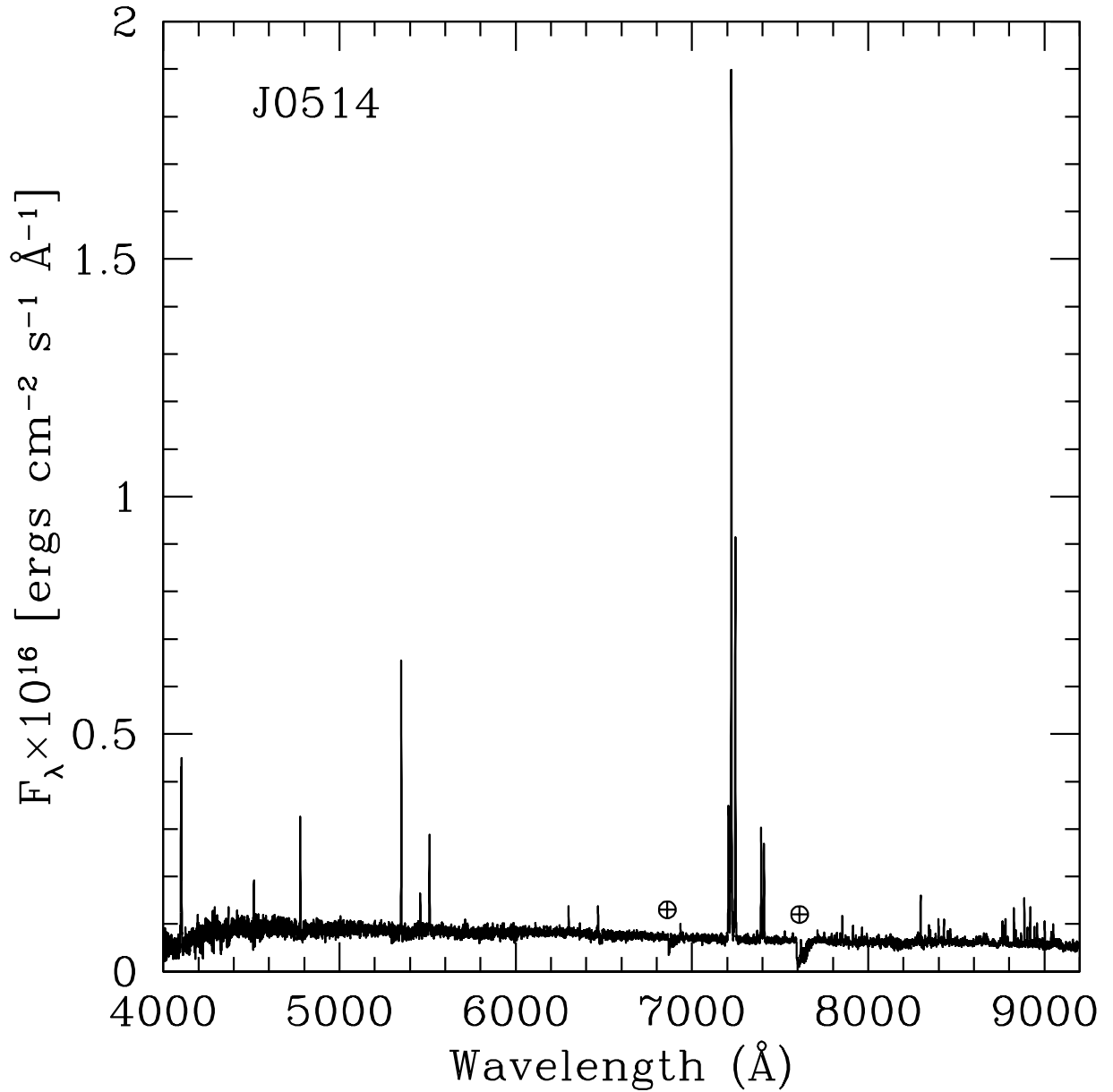}
\caption{\label{fig:elg} The spectrum of 2MASS~J05144238-6535497, a newly recognized emission line galaxy at z=0.101, and an example of an object identified as a photometric candidate for C\,{\sc ii} emission, but upon spectroscopy later proving to be a false positive, lacking C\,{\sc ii}. The reduced FITS-format spectrum of this, and all 19 objects in Table~2, are provided here as Data Behind Figure.}
 
\end{figure}

\appendix
\centerline{Interesting Objects Lacking C\,{\sc ii} Emission}
\bigskip
Although our narrowband imaging is an effective means of identifying C\,{\sc ii} emission objects, the technique inevitably also has false positives.  Further explanation and examples of this phenomenon are given in Paper~I, and details of these objects in the current sample are given in Table~2, together with suggested classifications based on our spectra and also photometry in the literature. The individual FITS-format spectra of all 19 objects in this category are provided as Data Behind Figure in Figure~\ref{fig:elg}. In some cases, the photometry in the literature makes it clear that the star is likely a Long Period Variable (LPV) or early-type eclipsing binary. The magnitudes in the table must thus be regarded with caution, as high amplitude photometric variations are known to be present in at least some cases.

For a handful of the objects, our spectra provide no obvious explanation of our photometrically measured C\,{\sc ii} excess; a few may be simple statistical fluctuations. Intrinsic stellar variability between the survey 300~s on-band and off-band images is also a possibility.

A few of the false positives prove to be interesting objects in their own right, even though they lack $\lambda\lambda$7231, 7236 emission. We discuss these briefly below.
\bigskip

{\it J05022531-6552527}: There is no previous spectral information in the literature. We classify this object as B[e], with quite weak but still well-detected nebular forbidden lines of  [O\,{\sc ii}], [N\,{\sc ii}], and [S\,{\sc ii}] in our spectrum. These stars are rare but not unknown in the Magellanic Clouds \citep{Zickgraf}.

\bigskip

{\it J05060423-7016513}: This object is cataloged as a carbon star, WORC~82 \citep{Westerlund78}, and a known high-amplitude ($\sim$2.5~mag) variable. \citet{Ogrady} give a period (840~d) and light curve. Our spectrum shows no obvious C$_2$ or CN bands, but unfortunately occurred near minimum light and is thus poorly exposed. Strong Balmer and Ca IR triplet emission, as well as the NaD doublet in absorption, are evident. Extremely prominent Li\,{\sc i} $\lambda$6707 absorption is also present, a typical characteristic of the highly variable ``super-AGB stars" discussed by \citet{Ogrady}.

\bigskip

 {\it J05144238-6535497}: This object proves to be a previously uncatalogued emission line galaxy (ELG) at z=0.101, with our spectrum (Figure~\ref{fig:elg}) showing strong, narrow Balmer, [O\,{\sc ii}], [O\,{\sc iii}], [N\,{\sc ii}], and [S\,{\sc ii}] emission. At this redshift, the H$\alpha$ and [N\,{\sc ii}] emission fall within our on-band C\,{\sc ii} $\lambda\lambda$7231, 7236 imaging filter, explaining the selection of the object in our sample. Paper~I notes a similar example, but in that case with a previously-known ELG, \hbox{6dFGS~gJ042936.9-692653} \citep{Jones09}, again at z=0.101, also selected as a [WC] candidate in our photometry. The large angular separation on the sky of these two ELG rules out any physical association; the redshift coincidence is purely due to observational selection. Multiple photometric surveys in the literature report quite different magnitudes for this object, and variability, if genuine, would be unusual. However, at this modest redshift, it is likely that the object is partially resolved by some but not all of the surveys, possibly explaining the observed scatter. The object appears stellar in our narrowband imaging. Both the position in the BPT diagram \citep{BPT} and the high FIR flux cataloged in the {\it Spitzer} IRAC4 band indicate a starburst galaxy. However the very low [S\,{\sc ii}]/$H\alpha$ ratio is at the extreme end of the \citet{Kewley} distribution and may be worthy of further study.

\bigskip

{\it J05313424-6901217}: We find no previous spectroscopy cited in the literature. Our spectrum is that of a Be shell star, with the classic H$\alpha$ profile of broad emission bisected by strong, narrow absorption, as well as broad He\,{\sc i} absorption.

\bigskip

 {\it J05334283-7003193 }:  We classify this object as an S star, confirming the tentative suggestion by \citet{Hughes}. There are multiple, prominent ZrO bands, NaD absorption, very strong, narrow Balmer emission, and probable CN bands. No obvious Li\,{\sc i} $\lambda$6707 is seen. \citet{Smith} type this object as a carbon star.

\bibliography{WC11}{}

\begin{thebibliography}{}
\expandafter\ifx\csname natexlab\endcsname\relax\def\natexlab#1{#1}\fi
\providecommand{\url}[1]{\href{#1}{#1}}
\providecommand{\dodoi}[1]{doi:~\href{http://doi.org/#1}{\nolinkurl{#1}}}
\providecommand{\doeprint}[1]{\href{http://ascl.net/#1}{\nolinkurl{http://ascl.net/#1}}}
\providecommand{\doarXiv}[1]{\href{https://arxiv.org/abs/#1}{\nolinkurl{https://arxiv.org/abs/#1}}}

\bibitem[{{Acker} \& {Neiner}(2003)}]{Acker}
{Acker}, A., \& {Neiner}, C. 2003, \aap, 403, 659,
  \dodoi{10.1051/0004-6361:20030391}

\bibitem[{{Aerts} {et~al.}(2010){Aerts}, {Christensen-Dalsgaard}, \&
  {Kurtz}}]{2010aste.book.....A}
{Aerts}, C., {Christensen-Dalsgaard}, J., \& {Kurtz}, D.~W. 2010,
  {Asteroseismology} (Berlin: Springer)

\bibitem[{{Baldwin} {et~al.}(1981){Baldwin}, {Phillips}, \& {Terlevich}}]{BPT}
{Baldwin}, J.~A., {Phillips}, M.~M., \& {Terlevich}, R. 1981, \pasp, 93, 5,
  \dodoi{10.1086/130766}

\bibitem[{{Cardelli} {et~al.}(1989){Cardelli}, {Clayton}, \& {Mathis}}]{CCM}
{Cardelli}, J.~A., {Clayton}, G.~C., \& {Mathis}, J.~S. 1989, \apj, 345, 245,
  \dodoi{10.1086/167900}

\bibitem[{{Crowther} {et~al.}(1998){Crowther}, {De Marco}, \&
  {Barlow}}]{Crowther}
{Crowther}, P.~A., {De Marco}, O., \& {Barlow}, M.~J. 1998, \mnras, 296, 367,
  \dodoi{10.1046/j.1365-8711.1998.01360.x}

\bibitem[{{De Marco} {et~al.}(1997){De Marco}, {Barlow}, \&
  {Storey}}]{DeMarcoI}
{De Marco}, O., {Barlow}, M.~J., \& {Storey}, P.~J. 1997, \mnras, 292, 86,
  \dodoi{10.1093/mnras/292.1.86}

\bibitem[{{De Marco} \& {Crowther}(1998)}]{DeMarcoII}
{De Marco}, O., \& {Crowther}, P.~A. 1998, \mnras, 296, 419,
  \dodoi{10.1046/j.1365-8711.1998.01379.x}

\bibitem[{{De Marco} \& {Soker}(2002)}]{demarco02}
{De Marco}, O., \& {Soker}, N. 2002, \pasp, 114, 602, \dodoi{10.1086/341691}

\bibitem[{{De Marco} {et~al.}(1998){De Marco}, {Storey}, \&
  {Barlow}}]{DeMarcoIII}
{De Marco}, O., {Storey}, P.~J., \& {Barlow}, M.~J. 1998, \mnras, 297, 999,
  \dodoi{10.1046/j.1365-8711.1998.297004999.x}

\bibitem[{{Dopita} {et~al.}(1994){Dopita}, {Vassiliadis}, {Meatheringham},
  {Ford}, {Bohlin}, {Wood}, {Stecher}, {Maran}, \& {Harrington}}]{Dopita}
{Dopita}, M.~A., {Vassiliadis}, E., {Meatheringham}, S.~J., {et~al.} 1994,
  \apj, 426, 150, \dodoi{10.1086/174050}

\bibitem[{{Gaia Collaboration} {et~al.}(2016){Gaia Collaboration}, {Prusti},
  {de Bruijne}, {Brown}, {Vallenari}, {Babusiaux}, {Bailer-Jones}, {Bastian},
  {Biermann}, {Evans}, {Eyer}, {Jansen}, {Jordi}, {Klioner}, {Lammers},
  {Lindegren}, {Luri}, {Mignard}, {Milligan}, {Panem}, {Poinsignon},
  {Pourbaix}, {Randich}, {Sarri}, {Sartoretti}, {Siddiqui}, {Soubiran},
  {Valette}, {van Leeuwen}, {Walton}, {Aerts}, {Arenou}, {Cropper}, {Drimmel},
  {H{\o}g}, {Katz}, {Lattanzi}, {O'Mullane}, {Grebel}, {Holland}, {Huc},
  {Passot}, {Bramante}, {Cacciari}, {Casta{\~n}eda}, {Chaoul}, {Cheek}, {De
  Angeli}, {Fabricius}, {Guerra}, {Hern{\'a}ndez}, {Jean-Antoine-Piccolo},
  {Masana}, {Messineo}, {Mowlavi}, {Nienartowicz}, {Ord{\'o}{\~n}ez-Blanco},
  {Panuzzo}, {Portell}, {Richards}, {Riello}, {Seabroke}, {Tanga},
  {Th{\'e}venin}, {Torra}, {Els}, {Gracia-Abril}, {Comoretto},
  {Garcia-Reinaldos}, {Lock}, {Mercier}, {Altmann}, {Andrae}, {Astraatmadja},
  {Bellas-Velidis}, {Benson}, {Berthier}, {Blomme}, {Busso}, {Carry},
  {Cellino}, {Clementini}, {Cowell}, {Creevey}, {Cuypers}, {Davidson}, {De
  Ridder}, {de Torres}, {Delchambre}, {Dell'Oro}, {Ducourant}, {Fr{\'e}mat},
  {Garc{\'\i}a-Torres}, {Gosset}, {Halbwachs}, {Hambly}, {Harrison}, {Hauser},
  {Hestroffer}, {Hodgkin}, {Huckle}, {Hutton}, {Jasniewicz}, {Jordan},
  {Kontizas}, {Korn}, {Lanzafame}, {Manteiga}, {Moitinho}, {Muinonen},
  {Osinde}, {Pancino}, {Pauwels}, {Petit}, {Recio-Blanco}, {Robin}, {Sarro},
  {Siopis}, {Smith}, {Smith}, {Sozzetti}, {Thuillot}, {van Reeven}, {Viala},
  {Abbas}, {Abreu Aramburu}, {Accart}, {Aguado}, {Allan}, {Allasia},
  {Altavilla}, {{\'A}lvarez}, {Alves}, {Anderson}, {Andrei}, {Anglada Varela},
  {Antiche}, {Antoja}, {Ant{\'o}n}, {Arcay}, {Atzei}, {Ayache}, {Bach},
  {Baker}, {Balaguer-N{\'u}{\~n}ez}, {Barache}, {Barata}, {Barbier}, {Barblan},
  {Baroni}, {Barrado y Navascu{\'e}s}, {Barros}, {Barstow}, {Becciani},
  {Bellazzini}, {Bellei}, {Bello Garc{\'\i}a}, {Belokurov}, {Bendjoya},
  {Berihuete}, {Bianchi}, {Bienaym{\'e}}, {Billebaud}, {Blagorodnova},
  {Blanco-Cuaresma}, {Boch}, {Bombrun}, {Borrachero}, {Bouquillon}, {Bourda},
  {Bouy}, {Bragaglia}, {Breddels}, {Brouillet}, {Br{\"u}semeister},
  {Bucciarelli}, {Budnik}, {Burgess}, {Burgon}, {Burlacu}, {Busonero}, {Buzzi},
  {Caffau}, {Cambras}, {Campbell}, {Cancelliere}, {Cantat-Gaudin}, {Carlucci},
  {Carrasco}, {Castellani}, {Charlot}, {Charnas}, {Charvet}, {Chassat},
  {Chiavassa}, {Clotet}, {Cocozza}, {Collins}, {Collins}, {Costigan}, {Crifo},
  {Cross}, {Crosta}, {Crowley}, {Dafonte}, {Damerdji}, {Dapergolas}, {David},
  {David}, {De Cat}, {de Felice}, {de Laverny}, {De Luise}, {De March}, {de
  Martino}, {de Souza}, {Debosscher}, {del Pozo}, {Delbo}, {Delgado},
  {Delgado}, {di Marco}, {Di Matteo}, {Diakite}, {Distefano}, {Dolding}, {Dos
  Anjos}, {Drazinos}, {Dur{\'a}n}, {Dzigan}, {Ecale}, {Edvardsson}, {Enke},
  {Erdmann}, {Escolar}, {Espina}, {Evans}, {Eynard Bontemps}, {Fabre},
  {Fabrizio}, {Faigler}, {Falc{\~a}o}, {Farr{\`a}s Casas}, {Faye}, {Federici},
  {Fedorets}, {Fern{\'a}ndez-Hern{\'a}ndez}, {Fernique}, {Fienga}, {Figueras},
  {Filippi}, {Findeisen}, {Fonti}, {Fouesneau}, {Fraile}, {Fraser}, {Fuchs},
  {Furnell}, {Gai}, {Galleti}, {Galluccio}, {Garabato}, {Garc{\'\i}a-Sedano},
  {Gar{\'e}}, {Garofalo}, {Garralda}, {Gavras}, {Gerssen}, {Geyer}, {Gilmore},
  {Girona}, {Giuffrida}, {Gomes}, {Gonz{\'a}lez-Marcos},
  {Gonz{\'a}lez-N{\'u}{\~n}ez}, {Gonz{\'a}lez-Vidal}, {Granvik}, {Guerrier},
  {Guillout}, {Guiraud}, {G{\'u}rpide}, {Guti{\'e}rrez-S{\'a}nchez}, {Guy},
  {Haigron}, {Hatzidimitriou}, {Haywood}, {Heiter}, {Helmi}, {Hobbs},
  {Hofmann}, {Holl}, {Holland}, {Hunt}, {Hypki}, {Icardi}, {Irwin}, {Jevardat
  de Fombelle}, {Jofr{\'e}}, {Jonker}, {Jorissen}, {Julbe}, {Karampelas},
  {Kochoska}, {Kohley}, {Kolenberg}, {Kontizas}, {Koposov}, {Kordopatis},
  {Koubsky}, {Kowalczyk}, {Krone-Martins}, {Kudryashova}, {Kull}, {Bachchan},
  {Lacoste-Seris}, {Lanza}, {Lavigne}, {Le Poncin-Lafitte}, {Lebreton},
  {Lebzelter}, {Leccia}, {Leclerc}, {Lecoeur-Taibi}, {Lemaitre}, {Lenhardt},
  {Leroux}, {Liao}, {Licata}, {Lindstr{\o}m}, {Lister}, {Livanou}, {Lobel},
  {L{\"o}ffler}, {L{\'o}pez}, {Lopez-Lozano}, {Lorenz}, {Loureiro},
  {MacDonald}, {Magalh{\~a}es Fernandes}, {Managau}, {Mann}, {Mantelet},
  {Marchal}, {Marchant}, {Marconi}, {Marie}, {Marinoni}, {Marrese},
  {Marschalk{\'o}}, {Marshall}, {Mart{\'\i}n-Fleitas}, {Martino}, {Mary},
  {Matijevi{\v{c}}}, {Mazeh}, {McMillan}, {Messina}, {Mestre}, {Michalik},
  {Millar}, {Miranda}, {Molina}, {Molinaro}, {Molinaro}, {Moln{\'a}r},
  {Moniez}, {Montegriffo}, {Monteiro}, {Mor}, {Mora}, {Morbidelli}, {Morel},
  {Morgenthaler}, {Morley}, {Morris}, {Mulone}, {Muraveva}, {Musella},
  {Narbonne}, {Nelemans}, {Nicastro}, {Noval}, {Ord{\'e}novic},
  {Ordieres-Mer{\'e}}, {Osborne}, {Pagani}, {Pagano}, {Pailler}, {Palacin},
  {Palaversa}, {Parsons}, {Paulsen}, {Pecoraro}, {Pedrosa}, {Pentik{\"a}inen},
  {Pereira}, {Pichon}, {Piersimoni}, {Pineau}, {Plachy}, {Plum}, {Poujoulet},
  {Pr{\v{s}}a}, {Pulone}, {Ragaini}, {Rago}, {Rambaux}, {Ramos-Lerate},
  {Ranalli}, {Rauw}, {Read}, {Regibo}, {Renk}, {Reyl{\'e}}, {Ribeiro},
  {Rimoldini}, {Ripepi}, {Riva}, {Rixon}, {Roelens}, {Romero-G{\'o}mez},
  {Rowell}, {Royer}, {Rudolph}, {Ruiz-Dern}, {Sadowski}, {Sagrist{\`a}
  Sell{\'e}s}, {Sahlmann}, {Salgado}, {Salguero}, {Sarasso}, {Savietto},
  {Schnorhk}, {Schultheis}, {Sciacca}, {Segol}, {Segovia}, {Segransan},
  {Serpell}, {Shih}, {Smareglia}, {Smart}, {Smith}, {Solano}, {Solitro},
  {Sordo}, {Soria Nieto}, {Souchay}, {Spagna}, {Spoto}, {Stampa}, {Steele},
  {Steidelm{\"u}ller}, {Stephenson}, {Stoev}, {Suess}, {S{\"u}veges}, {Surdej},
  {Szabados}, {Szegedi-Elek}, {Tapiador}, {Taris}, {Tauran}, {Taylor},
  {Teixeira}, {Terrett}, {Tingley}, {Trager}, {Turon}, {Ulla}, {Utrilla},
  {Valentini}, {van Elteren}, {Van Hemelryck}, {van Leeuwen}, {Varadi},
  {Vecchiato}, {Veljanoski}, {Via}, {Vicente}, {Vogt}, {Voss}, {Votruba},
  {Voutsinas}, {Walmsley}, {Weiler}, {Weingrill}, {Werner}, {Wevers},
  {Whitehead}, {Wyrzykowski}, {Yoldas}, {{\v{Z}}erjal}, {Zucker}, {Zurbach},
  {Zwitter}, {Alecu}, {Allen}, {Allende Prieto}, {Amorim},
  {Anglada-Escud{\'e}}, {Arsenijevic}, {Azaz}, {Balm}, {Beck}, {Bernstein},
  {Bigot}, {Bijaoui}, {Blasco}, {Bonfigli}, {Bono}, {Boudreault}, {Bressan},
  {Brown}, {Brunet}, {Bunclark}, {Buonanno}, {Butkevich}, {Carret}, {Carrion},
  {Chemin}, {Ch{\'e}reau}, {Corcione}, {Darmigny}, {de Boer}, {de Teodoro}, {de
  Zeeuw}, {Delle Luche}, {Domingues}, {Dubath}, {Fodor}, {Fr{\'e}zouls},
  {Fries}, {Fustes}, {Fyfe}, {Gallardo}, {Gallegos}, {Gardiol}, {Gebran},
  {Gomboc}, {G{\'o}mez}, {Grux}, {Gueguen}, {Heyrovsky}, {Hoar}, {Iannicola},
  {Isasi Parache}, {Janotto}, {Joliet}, {Jonckheere}, {Keil}, {Kim},
  {Klagyivik}, {Klar}, {Knude}, {Kochukhov}, {Kolka}, {Kos}, {Kutka}, {Lainey},
  {LeBouquin}, {Liu}, {Loreggia}, {Makarov}, {Marseille}, {Martayan},
  {Martinez-Rubi}, {Massart}, {Meynadier}, {Mignot}, {Munari}, {Nguyen},
  {Nordlander}, {Ocvirk}, {O'Flaherty}, {Olias Sanz}, {Ortiz}, {Osorio},
  {Oszkiewicz}, {Ouzounis}, {Palmer}, {Park}, {Pasquato}, {Peltzer}, {Peralta},
  {P{\'e}turaud}, {Pieniluoma}, {Pigozzi}, {Poels}, {Prat}, {Prod'homme},
  {Raison}, {Rebordao}, {Risquez}, {Rocca-Volmerange}, {Rosen}, {Ruiz-Fuertes},
  {Russo}, {Sembay}, {Serraller Vizcaino}, {Short}, {Siebert}, {Silva},
  {Sinachopoulos}, {Slezak}, {Soffel}, {Sosnowska}, {Strai{\v{z}}ys}, {ter
  Linden}, {Terrell}, {Theil}, {Tiede}, {Troisi}, {Tsalmantza}, {Tur},
  {Vaccari}, {Vachier}, {Valles}, {Van Hamme}, {Veltz}, {Virtanen}, {Wallut},
  {Wichmann}, {Wilkinson}, {Ziaeepour}, \& {Zschocke}}]{gaia1}
{Gaia Collaboration}, {Prusti}, T., {de Bruijne}, J.~H.~J., {et~al.} 2016,
  \aap, 595, A1, \dodoi{10.1051/0004-6361/201629272}

\bibitem[{{Gaia Collaboration} {et~al.}(2022){Gaia Collaboration}, {Vallenari},
  {Brown}, {Prusti}, {de Bruijne}, {Arenou}, {Babusiaux}, {Biermann},
  {Creevey}, {Ducourant}, {Evans}, {Eyer}, {Guerra}, {Hutton}, {Jordi},
  {Klioner}, {Lammers}, {Lindegren}, {Luri}, {Mignard}, {Panem}, {Pourbaix},
  {Randich}, {Sartoretti}, {Soubiran}, {Tanga}, {Walton}, {Bailer-Jones},
  {Bastian}, {Drimmel}, {Jansen}, {Katz}, {Lattanzi}, {van Leeuwen}, {Bakker},
  {Cacciari}, {Casta{\~n}eda}, {De Angeli}, {Fabricius}, {Fouesneau},
  {Fr{\'e}mat}, {Galluccio}, {Guerrier}, {Heiter}, {Masana}, {Messineo},
  {Mowlavi}, {Nicolas}, {Nienartowicz}, {Pailler}, {Panuzzo}, {Riclet}, {Roux},
  {Seabroke}, {Sordo{\o}rcit}, {Th{\'e}venin}, {Gracia-Abril}, {Portell},
  {Teyssier}, {Altmann}, {Andrae}, {Audard}, {Bellas-Velidis}, {Benson},
  {Berthier}, {Blomme}, {Burgess}, {Busonero}, {Busso}, {C{\'a}novas}, {Carry},
  {Cellino}, {Cheek}, {Clementini}, {Damerdji}, {Davidson}, {de Teodoro},
  {Nu{\~n}ez Campos}, {Delchambre}, {Dell'Oro}, {Esquej},
  {Fern{\'a}ndez-Hern{\'a}ndez}, {Fraile}, {Garabato}, {Garc{\'\i}a-Lario},
  {Gosset}, {Haigron}, {Halbwachs}, {Hambly}, {Harrison}, {Hern{\'a}ndez},
  {Hestroffer}, {Hodgkin}, {Holl}, {Jan{\ss}en}, {Jevardat de Fombelle},
  {Jordan}, {Krone-Martins}, {Lanzafame}, {L{\"o}ffler}, {Marchal}, {Marrese},
  {Moitinho}, {Muinonen}, {Osborne}, {Pancino}, {Pauwels}, {Recio-Blanco},
  {Reyl{\'e}}, {Riello}, {Rimoldini}, {Roegiers}, {Rybizki}, {Sarro}, {Siopis},
  {Smith}, {Sozzetti}, {Utrilla}, {van Leeuwen}, {Abbas}, {{\'A}brah{\'a}m},
  {Abreu Aramburu}, {Aerts}, {Aguado}, {Ajaj}, {Aldea-Montero}, {Altavilla},
  {{\'A}lvarez}, {Alves}, {Anders}, {Anderson}, {Anglada Varela}, {Antoja},
  {Baines}, {Baker}, {Balaguer-N{\'u}{\~n}ez}, {Balbinot}, {Balog}, {Barache},
  {Barbato}, {Barros}, {Barstow}, {Bartolom{\'e}}, {Bassilana}, {Bauchet},
  {Becciani}, {Bellazzini}, {Berihuete}, {Bernet}, {Bertone}, {Bianchi},
  {Binnenfeld}, {Blanco-Cuaresma}, {Blazere}, {Boch}, {Bombrun}, {Bossini},
  {Bouquillon}, {Bragaglia}, {Bramante}, {Breedt}, {Bressan}, {Brouillet},
  {Brugaletta}, {Bucciarelli}, {Burlacu}, {Butkevich}, {Buzzi}, {Caffau},
  {Cancelliere}, {Cantat-Gaudin}, {Carballo}, {Carlucci}, {Carnerero},
  {Carrasco}, {Casamiquela}, {Castellani}, {Castro-Ginard}, {Chaoul},
  {Charlot}, {Chemin}, {Chiaramida}, {Chiavassa}, {Chornay}, {Comoretto},
  {Contursi}, {Cooper}, {Cornez}, {Cowell}, {Crifo}, {Cropper}, {Crosta},
  {Crowley}, {Dafonte}, {Dapergolas}, {David}, {David}, {de Laverny}, {De
  Luise}, {De March}, {De Ridder}, {de Souza}, {de Torres}, {del Peloso}, {del
  Pozo}, {Delbo}, {Delgado}, {Delisle}, {Demouchy}, {Dharmawardena}, {Di
  Matteo}, {Diakite}, {Diener}, {Distefano}, {Dolding}, {Edvardsson}, {Enke},
  {Fabre}, {Fabrizio}, {Faigler}, {Fedorets}, {Fernique}, {Fienga}, {Figueras},
  {Fournier}, {Fouron}, {Fragkoudi}, {Gai}, {Garcia-Gutierrez},
  {Garcia-Reinaldos}, {Garc{\'\i}a-Torres}, {Garofalo}, {Gavel}, {Gavras},
  {Gerlach}, {Geyer}, {Giacobbe}, {Gilmore}, {Girona}, {Giuffrida}, {Gomel},
  {Gomez}, {Gonz{\'a}lez-N{\'u}{\~n}ez}, {Gonz{\'a}lez-Santamar{\'\i}a},
  {Gonz{\'a}lez-Vidal}, {Granvik}, {Guillout}, {Guiraud},
  {Guti{\'e}rrez-S{\'a}nchez}, {Guy}, {Hatzidimitriou}, {Hauser}, {Haywood},
  {Helmer}, {Helmi}, {Sarmiento}, {Hidalgo}, {Hilger}, {H{\l}adczuk}, {Hobbs},
  {Holland}, {Huckle}, {Jardine}, {Jasniewicz}, {Jean-Antoine Piccolo},
  {Jim{\'e}nez-Arranz}, {Jorissen}, {Juaristi Campillo}, {Julbe}, {Karbevska},
  {Kervella}, {Khanna}, {Kontizas}, {Kordopatis}, {Korn}, {K{\'o}sp{\'a}l},
  {Kostrzewa-Rutkowska}, {Kruszy{\'n}ska}, {Kun}, {Laizeau}, {Lambert},
  {Lanza}, {Lasne}, {Le Campion}, {Lebreton}, {Lebzelter}, {Leccia}, {Leclerc},
  {Lecoeur-Taibi}, {Liao}, {Licata}, {Lindstr{\o}m}, {Lister}, {Livanou},
  {Lobel}, {Lorca}, {Loup}, {Madrero Pardo}, {Magdaleno Romeo}, {Managau},
  {Mann}, {Manteiga}, {Marchant}, {Marconi}, {Marcos}, {Marcos Santos},
  {Mar{\'\i}n Pina}, {Marinoni}, {Marocco}, {Marshall}, {Polo},
  {Mart{\'\i}n-Fleitas}, {Marton}, {Mary}, {Masip}, {Massari},
  {Mastrobuono-Battisti}, {Mazeh}, {McMillan}, {Messina}, {Michalik}, {Millar},
  {Mints}, {Molina}, {Molinaro}, {Moln{\'a}r}, {Monari}, {Mongui{\'o}},
  {Montegriffo}, {Montero}, {Mor}, {Mora}, {Morbidelli}, {Morel}, {Morris},
  {Muraveva}, {Murphy}, {Musella}, {Nagy}, {Noval}, {Oca{\~n}a}, {Ogden},
  {Ordenovic}, {Osinde}, {Pagani}, {Pagano}, {Palaversa}, {Palicio},
  {Pallas-Quintela}, {Panahi}, {Payne-Wardenaar}, {Pe{\~n}alosa Esteller},
  {Penttil{\"a}}, {Pichon}, {Piersimoni}, {Pineau}, {Plachy}, {Plum}, {Poggio},
  {Pr{\v{s}}a}, {Pulone}, {Racero}, {Ragaini}, {Rainer}, {Raiteri}, {Rambaux},
  {Ramos}, {Ramos-Lerate}, {Re Fiorentin}, {Regibo}, {Richards}, {Rios Diaz},
  {Ripepi}, {Riva}, {Rix}, {Rixon}, {Robichon}, {Robin}, {Robin}, {Roelens},
  {Rogues}, {Rohrbasser}, {Romero-G{\'o}mez}, {Rowell}, {Royer}, {Ruz Mieres},
  {Rybicki}, {Sadowski}, {S{\'a}ez N{\'u}{\~n}ez}, {Sagrist{\`a} Sell{\'e}s},
  {Sahlmann}, {Salguero}, {Samaras}, {Sanchez Gimenez}, {Sanna},
  {Santove{\~n}a}, {Sarasso}, {Schultheis}, {Sciacca}, {Segol}, {Segovia},
  {S{\'e}gransan}, {Semeux}, {Shahaf}, {Siddiqui}, {Siebert}, {Siltala},
  {Silvelo}, {Slezak}, {Slezak}, {Smart}, {Snaith}, {Solano}, {Solitro},
  {Souami}, {Souchay}, {Spagna}, {Spina}, {Spoto}, {Steele},
  {Steidelm{\"u}ller}, {Stephenson}, {S{\"u}veges}, {Surdej}, {Szabados},
  {Szegedi-Elek}, {Taris}, {Taylo}, {Teixeira}, {Tolomei}, {Tonello}, {Torra},
  {Torra}, {Torralba Elipe}, {Trabucchi}, {Tsounis}, {Turon}, {Ulla}, {Unger},
  {Vaillant}, {van Dillen}, {van Reeven}, {Vanel}, {Vecchiato}, {Viala},
  {Vicente}, {Voutsinas}, {Weiler}, {Wevers}, {Wyrzykowski}, {Yoldas}, {Yvard},
  {Zhao}, {Zorec}, {Zucker}, \& {Zwitter}}]{gaia3}
{Gaia Collaboration}, {Vallenari}, A., {Brown}, A.~G.~A., {et~al.} 2022, arXiv
  e-prints, arXiv:2208.00211.
\newblock \doarXiv{2208.00211}

\bibitem[{{Hajduk} {et~al.}(2020){Hajduk}, {Todt}, {Hamann}, {Borek}, {van
  Hoof}, \& {Zijlstra}}]{Hajduk}
{Hajduk}, M., {Todt}, H., {Hamann}, W.-R., {et~al.} 2020, \mnras, 498, 1205,
  \dodoi{10.1093/mnras/staa2274}

\bibitem[{{Herald} \& {Bianchi}(2004)}]{Herald}
{Herald}, J.~E., \& {Bianchi}, L. 2004, \apj, 611, 294, \dodoi{10.1086/422133}

\bibitem[{{Herwig}(2001)}]{herwig}
{Herwig}, F. 2001, in Astrophysics and Space Science Library, Vol. 265,
  Astrophysics and Space Science Library, ed. R.~{Szczerba} \& S.~K.
  {G{\'o}rny}, 249, \dodoi{10.1007/978-94-015-9688-6_38}

\bibitem[{{Herwig} {et~al.}(1999){Herwig}, {Bl{\"o}cker}, {Langer}, \&
  {Driebe}}]{1999A&A...349L...5H}
{Herwig}, F., {Bl{\"o}cker}, T., {Langer}, N., \& {Driebe}, T. 1999, \aap, 349,
  L5.
\newblock \doarXiv{astro-ph/9908108}

\bibitem[{{Hillier} \& {Lanz}(2001)}]{Hillier01}
{Hillier}, D.~J., \& {Lanz}, T. 2001, in Astronomical Society of the Pacific
  Conference Series, Vol. 247, Spectroscopic Challenges of Photoionized
  Plasmas, ed. G.~{Ferland} \& D.~W. {Savin}, 343

\bibitem[{{Hillier} \& {Miller}(1998)}]{Hillier98}
{Hillier}, D.~J., \& {Miller}, D.~L. 1998, \apj, 496, 407,
  \dodoi{10.1086/305350}

\bibitem[{{Hrivnak} {et~al.}(2015){Hrivnak}, {Lu}, {Volk}, {Szczerba},
  {Soszy{\'n}ski}, \& {Hajduk}}]{hrivnak}
{Hrivnak}, B.~J., {Lu}, W., {Volk}, K., {et~al.} 2015, \apj, 805, 78,
  \dodoi{10.1088/0004-637X/805/1/78}

\bibitem[{{Hughes} \& {Wood}(1990)}]{Hughes}
{Hughes}, S. M.~G., \& {Wood}, P.~R. 1990, \aj, 99, 784, \dodoi{10.1086/115374}

\bibitem[{{Jacoby}(1980)}]{jacoby}
{Jacoby}, G.~H. 1980, \apjs, 42, 1, \dodoi{10.1086/190642}

\bibitem[{{Jones} {et~al.}(2009){Jones}, {Read}, {Saunders}, {Colless},
  {Jarrett}, {Parker}, {Fairall}, {Mauch}, {Sadler}, {Watson}, {Burton},
  {Campbell}, {Cass}, {Croom}, {Dawe}, {Fiegert}, {Frankcombe}, {Hartley},
  {Huchra}, {James}, {Kirby}, {Lahav}, {Lucey}, {Mamon}, {Moore}, {Peterson},
  {Prior}, {Proust}, {Russell}, {Safouris}, {Wakamatsu}, {Westra}, \&
  {Williams}}]{Jones09}
{Jones}, D.~H., {Read}, M.~A., {Saunders}, W., {et~al.} 2009, \mnras, 399, 683,
  \dodoi{10.1111/j.1365-2966.2009.15338.x}

\bibitem[{{Kewley} {et~al.}(2006){Kewley}, {Groves}, {Kauffmann}, \&
  {Heckman}}]{Kewley}
{Kewley}, L.~J., {Groves}, B., {Kauffmann}, G., \& {Heckman}, T. 2006, \mnras,
  372, 961, \dodoi{10.1111/j.1365-2966.2006.10859.x}

\bibitem[{{Kwitter} \& {Henry}(2022)}]{Kwitter}
{Kwitter}, K.~B., \& {Henry}, R.~B.~C. 2022, \pasp, 134, 022001,
  \dodoi{10.1088/1538-3873/ac32b1}

\bibitem[{{Leuenhagen} \& {Hamann}(1994)}]{Leu1}
{Leuenhagen}, U., \& {Hamann}, W.~R. 1994, \aap, 283, 567

\bibitem[{{Leuenhagen} \& {Hamann}(1998)}]{Leu3}
---. 1998, \aap, 330, 265

\bibitem[{{Leuenhagen} {et~al.}(1996){Leuenhagen}, {Hamann}, \&
  {Jeffery}}]{Leu2}
{Leuenhagen}, U., {Hamann}, W.~R., \& {Jeffery}, C.~S. 1996, \aap, 312, 167

\bibitem[{{Liu} {et~al.}(2000){Liu}, {Storey}, {Barlow}, {Danziger}, {Cohen},
  \& {Bryce}}]{Liu}
{Liu}, X.~W., {Storey}, P.~J., {Barlow}, M.~J., {et~al.} 2000, \mnras, 312,
  585, \dodoi{10.1046/j.1365-8711.2000.03167.x}

\bibitem[{{Margon} {et~al.}(2020{\natexlab{a}}){Margon}, {Manea}, {Williams},
  {Bond}, {Prochaska}, {Szyma{\'n}ski}, \& {Morrell}}]{Margon20a}
{Margon}, B., {Manea}, C., {Williams}, R., {et~al.} 2020{\natexlab{a}}, \apj,
  888, 54, \dodoi{10.3847/1538-4357/ab5e78}

\bibitem[{{Margon} {et~al.}(2021){Margon}, {Massey}, {Neugent}, \&
  {Morrell}}]{Margon21}
{Margon}, B., {Massey}, P., {Neugent}, K., \& {Morrell}, N. 2021, in American
  Astronomical Society Meeting Abstracts, Vol.~53, American Astronomical
  Society Meeting Abstracts, 548.10

\bibitem[{{Margon} {et~al.}(2020{\natexlab{b}}){Margon}, {Massey}, {Neugent},
  \& {Morrell}}]{Margon20b}
{Margon}, B., {Massey}, P., {Neugent}, K.~F., \& {Morrell}, N.
  2020{\natexlab{b}}, \apj, 898, 85 (Paper I), \dodoi{10.3847/1538-4357/ab9abe}

\bibitem[{{Mashburn} {et~al.}(2016){Mashburn}, {Sterling}, {Madonna},
  {Dinerstein}, {Roederer}, \& {Geballe}}]{Mashburn}
{Mashburn}, A.~L., {Sterling}, N.~C., {Madonna}, S., {et~al.} 2016, \apjl, 831,
  L3, \dodoi{10.3847/2041-8205/831/1/L3}

\bibitem[{{Massey} {et~al.}(2007){Massey}, {Olsen}, {Hodge}, {Jacoby},
  {McNeill}, {Smith}, \& {Strong}}]{LGGSII}
{Massey}, P., {Olsen}, K.~A.~G., {Hodge}, P.~W., {et~al.} 2007, \aj, 133, 2393,
  \dodoi{10.1086/513319}

\bibitem[{{Matsuura} {et~al.}(2014){Matsuura}, {Bernard-Salas}, {Lloyd Evans},
  {Volk}, {Hrivnak}, {Sloan}, {Chu}, {Gruendl}, {Kraemer}, {Peeters},
  {Szczerba}, {Wood}, {Zijlstra}, {Hony}, {Ita}, {Kamath}, {Lagadec}, {Parker},
  {Reid}, {Shimonishi}, {Van Winckel}, {Woods}, {Kemper}, {Meixner}, {Otsuka},
  {Sahai}, {Sargent}, {Hora}, \& {McDonald}}]{Mats}
{Matsuura}, M., {Bernard-Salas}, J., {Lloyd Evans}, T., {et~al.} 2014, \mnras,
  439, 1472, \dodoi{10.1093/mnras/stt2495}

\bibitem[{{Meatheringham} \& {Dopita}(1991)}]{Meat91}
{Meatheringham}, S.~J., \& {Dopita}, M.~A. 1991, \apjs, 75, 407,
  \dodoi{10.1086/191536}

\bibitem[{{Miller Bertolami}(2016)}]{2016A&A...588A..25M}
{Miller Bertolami}, M.~M. 2016, \aap, 588, A25,
  \dodoi{10.1051/0004-6361/201526577}

\bibitem[{{Monk} {et~al.}(1988){Monk}, {Barlow}, \& {Clegg}}]{Monk}
{Monk}, D.~J., {Barlow}, M.~J., \& {Clegg}, R.~E.~S. 1988, \mnras, 234, 583,
  \dodoi{10.1093/mnras/234.3.583}

\bibitem[{{O'Grady} {et~al.}(2020){O'Grady}, {Drout}, {Shappee}, {Bauer},
  {Fuller}, {Kochanek}, {Jayasinghe}, {Gaensler}, {Stanek}, {Holoien},
  {Prieto}, \& {Thompson}}]{Ogrady}
{O'Grady}, A. J.~G., {Drout}, M.~R., {Shappee}, B.~J., {et~al.} 2020, \apj,
  901, 135, \dodoi{10.3847/1538-4357/abafad}

\bibitem[{{Otsuka} {et~al.}(2017){Otsuka}, {Ueta}, {van Hoof}, {Sahai},
  {Aleman}, {Zijlstra}, {Chu}, {Villaver}, {Leal-Ferreira}, {Kastner},
  {Szczerba}, \& {Exter}}]{Otsuka}
{Otsuka}, M., {Ueta}, T., {van Hoof}, P. A.~M., {et~al.} 2017, \apjs, 231, 22,
  \dodoi{10.3847/1538-4365/aa8175}

\bibitem[{{Pe\~na} {et~al.}(1997){Pe\~na}, {Ruiz}, \& {Torres-Peimbert}}]{Pena}
{Pe\~na}, M., {Ruiz}, M.~T., \& {Torres-Peimbert}, S. 1997, \aap, 324, 674

\bibitem[{{Pietrzy{\'n}ski} {et~al.}(2019){Pietrzy{\'n}ski}, {Graczyk},
  {Gallenne}, {Gieren}, {Thompson}, {Pilecki}, {Karczmarek}, {G{\'o}rski},
  {Suchomska}, {Taormina}, {Zgirski}, {Wielg{\'o}rski}, {Ko{\l}aczkowski},
  {Konorski}, {Villanova}, {Nardetto}, {Kervella}, {Bresolin}, {Kudritzki},
  {Storm}, {Smolec}, \& {Narloch}}]{Piet}
{Pietrzy{\'n}ski}, G., {Graczyk}, D., {Gallenne}, A., {et~al.} 2019, \nat, 567,
  200, \dodoi{10.1038/s41586-019-0999-4}

\bibitem[{{Rate} \& {Crowther}(2020)}]{2020MNRAS.493.1512R}
{Rate}, G., \& {Crowther}, P.~A. 2020, \mnras, 493, 1512,
  \dodoi{10.1093/mnras/stz3614}

\bibitem[{{Richter} {et~al.}(1987){Richter}, {Tammann}, \&
  {Huchtmeier}}]{Richter}
{Richter}, O.~G., {Tammann}, G.~A., \& {Huchtmeier}, W.~K. 1987, \aap, 171, 33

\bibitem[{{Sanduleak} {et~al.}(1978){Sanduleak}, {MacConnell}, \&
  {Philip}}]{SMP}
{Sanduleak}, N., {MacConnell}, D.~J., \& {Philip}, A.~G.~D. 1978, \pasp, 90,
  621, \dodoi{10.1086/130397}

\bibitem[{{Schlegel} {et~al.}(1998){Schlegel}, {Finkbeiner}, \&
  {Davis}}]{Schlegel}
{Schlegel}, D.~J., {Finkbeiner}, D.~P., \& {Davis}, M. 1998, \apj, 500, 525,
  \dodoi{10.1086/305772}

\bibitem[{{Shaw} {et~al.}(2001){Shaw}, {Stanghellini}, {Mutchler}, {Balick}, \&
  {Blades}}]{Shaw}
{Shaw}, R.~A., {Stanghellini}, L., {Mutchler}, M., {Balick}, B., \& {Blades},
  J.~C. 2001, \apj, 548, 727, \dodoi{10.1086/319013}

\bibitem[{{Skrutskie} {et~al.}(2006){Skrutskie}, {Cutri}, {Stiening},
  {Weinberg}, {Schneider}, {Carpenter}, {Beichman}, {Capps}, {Chester},
  {Elias}, {Huchra}, {Liebert}, {Lonsdale}, {Monet}, {Price}, {Seitzer},
  {Jarrett}, {Kirkpatrick}, {Gizis}, {Howard}, {Evans}, {Fowler}, {Fullmer},
  {Hurt}, {Light}, {Kopan}, {Marsh}, {McCallon}, {Tam}, {Van Dyk}, \&
  {Wheelock}}]{2MASS}
{Skrutskie}, M.~F., {Cutri}, R.~M., {Stiening}, R., {et~al.} 2006, \aj, 131,
  1163, \dodoi{10.1086/498708}

\bibitem[{{Smith} {et~al.}(1995){Smith}, {Plez}, {Lambert}, \&
  {Lubowich}}]{Smith}
{Smith}, V.~V., {Plez}, B., {Lambert}, D.~L., \& {Lubowich}, D.~A. 1995, \apj,
  441, 735, \dodoi{10.1086/175395}

\bibitem[{{Stanghellini} {et~al.}(1999){Stanghellini}, {Blades}, {Osmer},
  {Barlow}, \& {Liu}}]{Stan}
{Stanghellini}, L., {Blades}, J.~C., {Osmer}, S.~J., {Barlow}, M.~J., \& {Liu},
  X.~W. 1999, \apj, 510, 687, \dodoi{10.1086/306602}

\bibitem[{{van Aarle} {et~al.}(2011){van Aarle}, {van Winckel}, {Lloyd Evans},
  {Ueta}, {Wood}, \& {Ginsburg}}]{vanaarle}
{van Aarle}, E., {van Winckel}, H., {Lloyd Evans}, T., {et~al.} 2011, \aap,
  530, A90, \dodoi{10.1051/0004-6361/201015834}

\bibitem[{{van der Hucht}(2001)}]{vanderHucht2001}
{van der Hucht}, K.~A. 2001, \nar, 45, 135,
  \dodoi{10.1016/S1387-6473(00)00112-3}

\bibitem[{{van der Hucht} {et~al.}(1981){van der Hucht}, {Conti}, {Lundstrom},
  \& {Stenholm}}]{vanderHucht81}
{van der Hucht}, K.~A., {Conti}, P.~S., {Lundstrom}, I., \& {Stenholm}, B.
  1981, \ssr, 28, 227, \dodoi{10.1007/BF00173260}

\bibitem[{{Webster} \& {Glass}(1974)}]{Webster}
{Webster}, B.~L., \& {Glass}, I.~S. 1974, \mnras, 166, 491,
  \dodoi{10.1093/mnras/166.2.491}

\bibitem[{{Westerlund} {et~al.}(1978){Westerlund}, {Olander}, {Richer}, \&
  {Crabtree}}]{Westerlund78}
{Westerlund}, B.~E., {Olander}, N., {Richer}, H.~B., \& {Crabtree}, D.~R. 1978,
  \aaps, 31, 61

\bibitem[{{Westerlund} \& {Smith}(1964)}]{Westerlund64}
{Westerlund}, B.~E., \& {Smith}, L.~F. 1964, \mnras, 127, 449,
  \dodoi{10.1093/mnras/127.5.449}

\bibitem[{{Williams} {et~al.}(2021){Williams}, {Manea}, {Margon}, \&
  {Morrell}}]{Williams}
{Williams}, R., {Manea}, C., {Margon}, B., \& {Morrell}, N. 2021, \apj, 906,
  31, \dodoi{10.3847/1538-4357/abc754}

\bibitem[{{Zaritsky} {et~al.}(2004){Zaritsky}, {Harris}, {Thompson}, \&
  {Grebel}}]{zaritsky}
{Zaritsky}, D., {Harris}, J., {Thompson}, I.~B., \& {Grebel}, E.~K. 2004, \aj,
  128, 1606, \dodoi{10.1086/423910}

\bibitem[{{Zhang} {et~al.}(2005){Zhang}, {Liu}, {Luo}, {P{\'e}quignot}, \&
  {Barlow}}]{Zhang}
{Zhang}, Y., {Liu}, X.~W., {Luo}, S.~G., {P{\'e}quignot}, D., \& {Barlow},
  M.~J. 2005, \aap, 442, 249, \dodoi{10.1051/0004-6361:20052869}

\bibitem[{{Zickgraf}(2006)}]{Zickgraf}
{Zickgraf}, F.~J. 2006, in Astronomical Society of the Pacific Conference
  Series, Vol. 355, Stars with the B[e] Phenomenon, ed. M.~{Kraus} \& A.~S.
  {Miroshnichenko}, 135

\end{thebibliography}
\bibliographystyle{aasjournal}

\clearpage

\clearpage

\clearpage
\begin{deluxetable}{ccccccccc}[ht]
\tabletypesize{\small}
\tablecaption{Spectroscopically Verified LMC C\,{\sc ii} Emission Stars \label{tab:yesc2}}
 
\tablehead{
\colhead{Star (2MASS J+)}
 &\colhead{$V$}
 &\colhead{$M_V$}
 &\colhead{$(B-V)$}
 &\colhead{$(U-B)$}
 &\colhead{$\lambda7231$/$H\beta$}
 &\colhead{$\lambda7231$/$\lambda7236$}
 &\colhead{Class.}
 & Ref.\\
 }
\startdata
04383478-7036434  & 15.05 & -3.8 & 1.04 & -0.35 & 0.002 & 0.52 & [WC4] & 1 \\
05073893-6826061 &  15.65 & -3.2 & -0.03 & -1.09 & ... & 1.0 & ? & 2 \\
05112369-7001573  & 15.01 & -3.9 & 1.21 & -0.25 & 0.004 & 0.48 & [WC5] & 1 \\
05205243-7009354   & 16.33 & -2.5 & 1.22 & 0.40 & 0.006 & 0.52 & [WC4-5] & 1 \\
05242076-7005015   & 15.37 &  -3.5 & 0.87 & -0.30 & 0.004 & 0.51 & [WC5] & 2 \\
05312172-7017394 & 16.27 & -2.6 & -0.16 & -0.75 & 0.90 & 0.77 & [WC11] & 2 \\
05403079-6617374 &  15.71 & -3.2 & 0.82 & -0.70  & 0.004 & 0.47 & [WC8] & 1 \\
05582596-6944257 &  15.42 & -3.5 & 0.63 & -0.07 & 2.8 & 1.0 & [WC11] & 1, 3 \\
06081992-7157373 &  15.64 & -3.2 & 0.02 & -0.76 & 2.6 & 0.88 & [WC11] & 4, 5 \\
\enddata
\tablecomments{Here and in Table~2, the photometry is from \citet{zaritsky}. See \S3.1 for assumptions on distance and extinction used to derive $M_V$. References: (1) this work (2) Paper~I \hbox{(3) \citet{vanaarle}} (4)  \citet{Margon20a} (5) \citet{Williams}  
}
\end{deluxetable}

\begin{deluxetable}{ccc}[ht]
\tabletypesize{\small}
\tablecaption{Spectra of Candidates Lacking C\,{\sc ii} Emission \label{tab:noc2}}
 
\tablehead{
\colhead{Star (2MASS J+)}
 &\colhead{$V$}
 &\colhead{Classification}\\
 }
\startdata
04423895-7213163 &   15.67 & M, H em. \\
04463516-7138149 &   16.52 & M, strong H em.\\  
04515565-6736030 &   15.49 & B \\
04564493-6830586 &   15.84 & M \\
05004581-6630309 &   15.80 & B\\
05022531-6552527 &   15.37 & B[e], forbidden em.\tablenotemark{*} \\
05060423-7016513 &   17.44 & strong Li\tablenotemark{*} \\
05103465-6838514 &   15.26 & early B \\
05144238-6535497 &   18.84 & ELG\tablenotemark{*} \\
05161783-6751129 &   18.15  & early M?, H em. \\
05200703-6939424 &   14.39 & Be\\
05225894-6733112 &   15.30 & M, strong H em.\\
05243527-7047096 &   16.58 & M, strong H em. \\
05250191-6831579 &   14.09 & O \\
05302828-7059236 &   13.60 & early B \\
05305652-6901279 &   14.53 & early B \\
05313424-6901217 &   15.26 & Be, shell star\tablenotemark{*} \\
05322576-7055473 &   15.55 & early B \\
05334283-7003193 &   17.09 & S star\tablenotemark{*} \\
\enddata
\tablenotetext{*}{see Appendix for further details}
\end{deluxetable}

\end{document}